\documentclass[12pt]{article}

\usepackage{amsmath,amssymb}

\newcommand{\be}{\begin{equation}}
\newcommand{\ee}{\end{equation}}

\def\tsum{\textstyle \sum }
\def\cN{{\cal N}}
\def\bea{\begin{eqnarray}}
\def\eea{\end{eqnarray}}
\def\nn{\nonumber}
\def\ui{{\underline{i}}}
\def\uj{{\underline{j}}}

\begin{document}
\begin{titlepage}

\begin{center}
{\Large\bf $\cN=3$ Superparticle Model} \vspace{1cm}

{\large\bf I.L. Buchbinder$\,{}^{a}$,
I.B. Samsonov$\,{}^{b}$}
\\[8pt]
\it\small a) Dept.\ of Theoretical Physics, Tomsk State
Pedagogical University, 634041 Tomsk, Russia
\\
{\tt joseph@tspu.edu.ru}\\[8pt]
b) Institut f\"ur Theoretische Physik, Leibniz Universit\"at
Hannover, 30167 Hannover, Germany \footnote{Alexander von Humboldt fellow at Leibniz Universit\"at Hannover.}
\\
$\&$ \\
Laboratory of Mathematical Physics, Tomsk Polytechnic University,
634050 Tomsk, Russia\\
{\tt samsonov@mph.phtd.tpu.edu.ru}
\end{center}
\vspace{0.5cm}

\begin{abstract}
\noindent We consider the formulation and quantization of the
$\cN=3$ superparticle model, both with and without central charge.
Without the central charge the action possesses $U(3)$ invariance
and therefore is naturally quantized in the $\cN=3$ harmonic
superspace. The quantization reproduces the $\cN=3$ supergauge
strength multiplets, described by analytic $\cN=3$ superfields and a
gravitino multiplet as a constrained $\cN=3$ chiral superfield. When
the central charge is present, it breaks the $U(3)$ R-symmetry of
$\cN=3$ superalgebra down to $SU(2)\times U(1)$, and the
corresponding superparticle model is formulated in the $\cN=2$
harmonic superspace extended by a pair of extra Grassmann variables.
The quantization of such a model leads to the massive BPS $\cN=3$
vector multiplet. It is shown that upon additional superfield
constraints such multiplet reduces to the massive $\cN=2$ vector
multiplet.
\end{abstract}
\end{titlepage}

\setcounter{equation}{0}
\section{Introduction}
The models of relativistic particles and superparticles have deep
relations to string and field theories. They can be considered not
only as toy models which hint how to quantize superstring theories,
but also describe the dynamics of D0-branes, point-like objects
which form a part of the physical content of the type IIA string
theory. From the quantum field theory point of view, the
quantization of superparticles results in superfield realizations of
supersymmetry multiplets with corresponding equations of motion and
constraints \cite{Casalbuoni,VolkovPashnev,BS,Luk1,Lus,Uvarov}. This
is especially important for the models with extended supersymmetry
since their superfield equations of motion are usually entangled
with superfield constraints \cite{N2constr}. In particular, the
unconstrained superfield formulation of $\cN=2$ supersymmetric
models of hypermultiplets and gauge multiplet are given within the
so-called harmonic superspace approach \cite{HSS,Book}. This
approach is crucial for the superfield quantization of these models
which usually requires the use of unconstrained superfields
\cite{HSS1} (see also \cite{Echaya} for applications to problem of
effective action).

We point out that free equations of motion for the $\cN=2$ super
Yang-Mills (SYM) and hypermultiplet models can be naturally
derived by quantizing the $\cN=2$ superparticle in harmonic
superspace \cite{Sorokin1,Sorokin2}.\footnote{The harmonic
superspace approach is also very effective for studying 1d
supersymmetric models with extended supersymmetry (see e.g., the
recent  works \cite{mech} and reviews \cite{mech1} on the
supersymmetric mechanics).} However, the unconstrained superfield
formulation for $\cN=4$ SYM theory has not been achieved yet
despite the many attempts made in this direction The $\cN=3$ SYM
theory in harmonic superspace \cite{Harm1}, which is known as a
maximally supersymmetric field theory with the unconstrained
superfield formulation, deserves special attention. Since the
$\cN=3$ SYM model is equivalent to the $\cN=4$ one on-shell, it
can be considered as an $\cN=3$ superfield formulation for the
$\cN=4$ SYM model. Some quantum aspects of the $\cN=3$ SYM model
in harmonic superspace were studied in \cite{Delduc,BISZ}.

Inspired by the success of harmonic superspace formulation for the
$\cN=3$ SYM model we pose the question: which $\cN=3$
supermultiplets, apart from supergauge one, admit the description
in terms of the $\cN=3$ superfields? For this purpose, we study
the relativistic superparticle model in the $\cN=3$ $d=4$ harmonic
superspace and quantize it. A generic discussion of $d=4$
superparticles in harmonic superspaces with $\cN\geq 2$ and the
existence in their quantum spectra of supergauge and supergravity
multiplets was given in \cite{Sorokin1,Sorokin2,Grassi}.

Note that different models of superparticles can be considered,
depending on whether their actions contain mass and central charge
terms. In this paper we consider in detail all such $\cN=3$
superparticles and find the superfield realizations of corresponding
supermultiplets. In particular, the Gupta-Bleuler quantization of
the massless $\cN=3$ superparticle without central charge reproduces
the $\cN=3$ SYM supermultiplet realized on superfield strengths.
These $\cN=3$ superfields satisfy the Grassmann and harmonic
shortness conditions \cite{Sorokin1,Ferrara}. Another interesting
multiplet appearing in this case is the $\cN=3$ gravitino multiplet
(with the highest helicity 3/2) which is described by a chiral
$\cN=3$ superfield.

The quantization of the $\cN=3$ superparticle with the central
charge term is not so straightforward. One of the features in this
case is that the central charge in the $\cN=3$ superalgebra breaks
the group of internal automorphisms $U(3)$ down to $SU(2)\times
U(1)$, and therefore the $SU(3)$ harmonic superspace approach is not
applicable here. Regarding the preserved R-symmetry group one can
use $SU(2)$ harmonics instead. Therefore the appropriate formulation
of such a superparticle is given in the $\cN=2$ harmonic superspace,
which is extended by a pair of additional Grassmann coordinates. We
show that the central charge term in this model coincides with the
central charge of $\cN=2$ harmonic superparticle studied in
\cite{Sorokin1,Sorokin2}. Hence, the quantization proceeds in the
same way as in the $\cN=2$ superparticle model and leads to the
supermultiplets of the $\cN=3$ supersymmetry with central charge,
realized on superfields in the $\cN=3$ superspace with $SU(2)$
harmonic variables.

One of the simplest multiplets appearing in the quantization of the
$\cN=2$ superparticle with a central charge is the massive
$q^+$-hypermultiplet described by an unconstrained $\cN=2$ analytic
superfield in harmonic superspace \cite{Sorokin1,Sorokin2}. In our
case, the quantization results in a similar $q^+$ superfield, which
depends on the extra Grassmann spinor coordinate in a chiral way.
Such a superfield describes the massive $\cN=3$ vector multiplet
where the mass is related to the central charge of the superalgebra
by the BPS condition. The on-shell field content of this multiplet
is given by 5 complex scalars, 4 Dirac spinors and 1 complex vector.
Thus it has twice as much components compared to the $\cN=2$ non-BPS
massive vector multiplet \cite{FS}.

Naturally, one is led to question, whether it is possible to impose
such extra constraints on the $\cN=3$ massive vector superfield,
which eliminate half of the states and reduce the above multiplet to
the $\cN=2$ massive vector multiplet. We give the positive answer to
this question and show that in $\cN=3$ superspace these constraints
look very similar to the equations of motion in the massive $\cN=1$
Wess-Zumino model. These equations preserve $\cN=3$ supersymmetry,
but violate CPT invariance of the multiplet. The resulting $\cN=3$
superfield describes exactly the $\cN=2$ massive vector multiplet
on-shell. In other words, we show that the massive non-BPS $\cN=2$
vector supermultiplet can be described by the $\cN=3$ superfield
under specific constraints, which violate the CPT invariance of the
$\cN=3$ superalgebra with central charge.

The paper is organized as follows. In Section 2 we introduce the
actions for the $\cN=3$ superparticle both with and without the
central charge term. The quantization of the massless $\cN=3$
superparticle without central charge term is considered in Section
3, where the $\cN=3$ gravitino and SYM multiplets are derived. In
Section 4 we quantize the $\cN=3$ superparticle with the central
charge and, in the simplest case, obtain the $\cN=3$ massive vector
supermultiplet. We also show that this multiplet can be reduced to
the massive $\cN=2$ vector supermultiplet by imposing extra
superfield constraints. In the Conclusion we summarize the results
and discuss some unresolved problems.

\section{$\cN=3$ superparticle action in harmonic superspace}
\subsection{$\cN=3$ superparticle model without central charge term}
The $\cN=3$ superspace is parameterized by coordinates
$Z^M=\{x^m,\theta_i^\alpha,\bar\theta^{i\dot\alpha}\}$, where
$\alpha,\dot\alpha,\ldots$ are the indices of $SL(2,C)$ and
$i,j,\ldots$ denote the indices of $SU(3)$. In this coordinate
system the superinvariant Cartan forms $\omega^M=\dot\omega^M d\tau$
are \be \omega^M=\left\{
\begin{array}{rcl}
\omega^m&=&dx^m-id\theta^\alpha_i\sigma^m_{\alpha\dot\alpha}
  \bar\theta^{i\dot\alpha}+i\theta_i^\alpha\sigma^m_{\alpha\dot\alpha}
   d\bar\theta^{i\dot\alpha}\\
\omega^\alpha_i&=&d\theta_i^\alpha\\
\bar\omega^{i\dot\alpha}&=&d\bar\theta^{i\dot\alpha}.
\end{array}
\right.
\label{e2}
\ee
In terms of the Cartan forms (\ref{e2}), the
massive superparticle action is given by \be S_{sp}=-\frac12\int
d\tau(e^{-1} \dot\omega^m\dot\omega_m+e m^2), \label{S1} \ee where
$e(\tau)$ is the einbein field, $m$ is the mass of the superparticle
and $\tau$ is the worldline parameter.

Action (\ref{S1}) is invariant under supertranslations
\bea
\delta_\epsilon \theta_i^\alpha&=&\epsilon_i^\alpha,\qquad
\delta_\epsilon \bar\theta^{i\dot\alpha}=
 \bar\epsilon^{i\dot\alpha},\nn\\
\delta_\epsilon x^m&=&-i\epsilon_i\sigma^m\bar\theta^i
+i\theta_i\sigma^m\bar\epsilon^i, \label{susy} \eea which correspond
to conserved charges (supercharges) \be
Q^i_\alpha=2ie^{-1}\dot\omega_m(\sigma^m\bar\theta^i)_\alpha,\qquad
\bar
Q_{i\dot\alpha}=-2ie^{-1}\dot\omega_m(\theta_i\sigma^m)_{\dot\alpha}.
\label{supercharges}
\ee
Together with the particle momenta, they
generate the $\cN=3$ superalgebra (\ref{N3-algebra}) after
quantization. The model (\ref{S1}) also respects the $U(3)$
R-symmetry of the $\cN=3$ superalgebra.

In general, the arbitrary superfields on the full $\cN=3$ superspace
with coordinates $Z^M$ (as well as the arbitrary superfields on any
extended superspace) have a large number of components and do not
correspond to irreducible representations of the $\cN=3$
superalgebra. The construction of the irreducible superfields with
fewer number of components is realized within the harmonic
superspace formalism \cite{HSS,Book}. Following this approach, in
the case of the $\cN=3$ supersymmetry \cite{Harm1}, one extends the
superspace $Z^M$ with the harmonic variables $u^I_i$ which are
$SU(3)$ matrices, \be
 u^\dag u={\bf 1}_{3\times3},\qquad \det u=1.
\label{su3}
\ee
Conditions (\ref{su3}) can be written in terms of
the matrix elements $u^I_i$ as \bea &&u^I_i\bar
u^i_J=\delta^I_J,\qquad u^I_i\bar u^j_I=\delta_i^j,
\label{e6}\\
&&
 \varepsilon^{ijk}u^1_i u^2_j u^3_k=1,\qquad
 \varepsilon_{ijk}\bar u_1^i \bar u_2^j \bar u_3^k=1.
\label{e7}
\eea
Here the capital Latin indices $I,J,\ldots$ are $SU(3)$ ones with
the values $1,2,3$. The invariant Cartan forms on the $SU(3)$ group
\cite{Sorokin1}
\be
\omega^I_J=du^I_i\bar u_J^i=-u^I_i d\bar u^i_J
\label{e12}
\ee
satisfy the following identity
\be
\omega^1_1+\omega^2_2+\omega^3_3=0,
\label{e14}
\ee
which can be derived from (\ref{e7}).

The action for a particle on the coset space $SU(3)/ (U(1)\times
U(1))$ can be constructed with the use of the Cartan forms
(\ref{e12}), \bea S_{SU(3)}&=&S_\omega+S_{WZ}+S_\lambda,
\label{e25}\\
S_\omega&=&\frac{R^2}2\int d\tau\, e^{-1}
 (\dot\omega^1_2\dot\omega^2_1+\dot\omega^2_3\dot\omega^3_2
 +\dot\omega^1_3\dot\omega^3_1),
\label{e26}\\
S_{WZ}&=&-\frac{is_1}2\int d\tau(\dot u^1_i\bar u_1^i- u^1_i\dot{\bar u}{}_1^i)
+\frac{is_2}2\int d\tau(\dot u^3_i \bar u_3^i-u^3_i\dot{\bar
u}{}_3^i),
\label{e27}\\
S_{\lambda}&=&\int d\tau\left[\tsum\limits_{I,J=1}^3
\lambda_I^J(u^I_i\bar u_J^i-\delta^I_J) +
\Lambda(\varepsilon^{ijk}u^1_i u^2_j u^3_k + \varepsilon_{ijk}\bar
u_1^i \bar u_2^j \bar u_3^k-2) \right].
\label{e28}
\eea
Action
$S_\omega$ is a kinetic term which appears here with the constant
$R^2/2$, $S_{WZ}$ is the Wess-Zumino term for harmonic variables
with the constants $s_1$, $s_2$, and action $S_\lambda$ takes into
account the constraints (\ref{e6},\ref{e7}) with the help of
Lagrange multipliers $(\lambda^I_J)=(\lambda^I_J)^\dag$ and
$\Lambda=\bar\Lambda$. The Lagrange multipliers have ten independent
real degrees of freedom and the corresponding constraints thus
single out eight independent components from eighteen components of
the arbitrary complex $3\times3$ matrix $u^I_i$.
In addition, there is the local $U(1)\times U(1)$ symmetry which
further reduces this number to six, i.e., the particle moves
effectively in the coset space $SU(3)/(U(1)\times U(1))$.

The variation of action (\ref{e28}) over the Lagrange multipliers
gives the following constraints for harmonic variables
\bea
\chi^I_J&\equiv&\frac{\delta S_{\lambda}}{\delta \lambda^J_I}
=u^I_i\bar u_J^i-\delta^I_J=0, \label{h-constr1}
\\
\chi_1&\equiv&\frac{\delta S_\lambda}{\delta\Lambda}
=\varepsilon^{ijk}u^1_iu^2_ju^3_k
+ \varepsilon_{ijk}\bar u_1^i\bar u_2^j\bar u_3^k-2=0.
\label{h-constr2}
\eea
Equation (\ref{h-constr1}) shows that the matrix $u$ is
unitary while (\ref{h-constr2}) means
${\rm Re}\det u=1$. These constraints uniquely imply
$\det u=1$. Therefore both conditions (\ref{e6},\ref{e7}) are
satisfied and the matrices $u^I_i$, $\bar u_I^i$
belong to the $SU(3)$ group. In particular,
for any $SU(3)$ matrix we have ${\rm Im}\det u=0$,
or,
\be
\chi_2
=\varepsilon^{ijk}u^1_iu^2_ju^3_k
- \varepsilon_{ijk}\bar u_1^i\bar u_2^j\bar u_3^k=0.
\label{h-constr3}
\ee
Hence, (\ref{h-constr3}) appears as a consequence of
(\ref{h-constr1},\ref{h-constr2}). This constraint will be used
in the next section for constructing the Dirac bracket.

The unitary matrices $u^I_i$ rotate the Grassmann variables and
supersymmetric Cartan forms (\ref{e2}),
\bea
\theta_i^\alpha&\to&\theta_I^\alpha=\bar u^i_I \theta_i^\alpha,\quad
\bar\theta^{i\dot\alpha}\to\bar\theta^{I\dot\alpha}=
u^I_i\bar\theta^{i\dot\alpha},\nonumber\\
\omega_i^\alpha&\to&\omega_I^\alpha=\bar
u_I^i\omega_i^\alpha,\quad
\bar\omega^{i\dot\alpha}\to \bar\omega^{I\dot\alpha}=
u^I_i\bar\omega^{i\dot\alpha}.
\label{e15}
\eea
The $\cN=3$ harmonic superspace is parameterized by a set of
coordinates $Z_H=\{x^m,\theta^\alpha_I,\bar\theta^{I\dot\alpha},u
\}$, where $\theta^\alpha_I$, $\bar\theta^{I\dot\alpha}$ are given by
(\ref{e15}). Apart from the usual complex conjugation, there is also
$\widetilde{\phantom{a}}$ conjugation which acts on the Grassmann
variables and harmonics as (cf. \cite{Book,Ferrara})
\bea
&&u^1_i\stackrel{\sim}{\leftrightarrow} \bar u_3^i,\quad
u^2_i\stackrel{\sim}{\leftrightarrow} -\bar u_2^i,\quad
u^3_i\stackrel{\sim}{\leftrightarrow} \bar u_1^i,
\label{e16}\nonumber
\\&&
\theta_1^\alpha \stackrel{\sim}{\leftrightarrow}
\bar\theta^{3\dot\alpha},\quad \theta_2^\alpha
\stackrel{\sim}{\leftrightarrow} -\bar\theta^{2\dot\alpha},\quad
\theta_3^\alpha \stackrel{\sim}{\leftrightarrow}
\bar\theta^{1\dot\alpha}.
\label{e17}
\eea
This conjugation is
natural in harmonic superspace $Z_H$, since the $\cN=3$ SYM action
is known to be real under (\ref{e17}). Applying the conjugation
$\widetilde{\phantom{a}}$ to action (\ref{e25}) swaps the constants
$s_1$, $s_2$ in the Wess-Zumino term. Therefore, action (\ref{e25})
is real under (\ref{e16}) if $s_1=s_2$.

The action of the $\cN=3$ superparticle without central charges
moving in the harmonic superspace $Z_H$ is a sum of (\ref{S1}) and
(\ref{e25}),
\be
S_{sp}+S_{SU(3)}=S_{sp}+S_\omega+S_{WZ}+S_\lambda
=\int d\tau\, L_1,
\label{e29}
\ee
where $L_1$ denotes the Lagrangian of the superparticle.

\subsection{$\cN=3$ superparticle model with the central charge term}
The superparticle action (\ref{S1}) admits the extension by the
Wess-Zumino term \cite{Luk1}, \be S_c=-\int d\tau(\bar
Z^{ij}\theta_i^\alpha\dot\theta_{j\alpha}
+Z_{ij}\bar\theta^i_{\dot\alpha}\dot{\bar\theta}{}^{j\dot\alpha}).
\label{S2} \ee Here, $Z_{ij}$ and its conjugate $\bar Z^{ij}$ are
the constant antisymmetric matrices, \be \bar Z^{ij}=-\bar
Z^{ji},\quad Z_{ij}=-Z_{ji},\quad (Z_{ij})^*=\bar Z^{ij}.
\label{e19}
\ee
The Wess--Zumino term (\ref{S2}) is also invariant
under supersymmetry (\ref{susy}), but up to a total derivative.
Added to action (\ref{S1}), it leads (upon taking into account
boundary contributions) to the conserved Noether supercharges with
central charge terms, \be
Q^i_\alpha=2ie^{-1}\dot\omega_m(\sigma^m\bar\theta^i)_\alpha
 +2\bar Z^{ij}\theta_{j\alpha},\qquad
\bar
Q_{i\dot\alpha}=-2ie^{-1}\dot\omega_m(\theta_i\sigma^m)_{\dot\alpha}
+2Z_{ij}\bar\theta^j_{\dot\alpha},
\label{supercharges1}
\ee
which
generate the $\cN=3$ superalgebra with central charge
(\ref{N3-algebra1}) after quantization. Therefore we refer to action
(\ref{S2}) as a central charge term in the superparticle model.

Since $Z_{ij}$ are constants, they break the $U(3)$ R-symmetry of
the $\cN=3$ superalgebra. To understand which symmetry survives, we
notice that any $3\times 3$ antisymmetric matrix is degenerate, \be
\det Z=0,\qquad \det \bar Z=0. \label{e20} \ee Moreover, performing
some rotation with the $SU(3)$ matrix $v^i_j$, the matrices
$Z_{ij}$, $\bar Z^{ij}$ can be brought to the normal form
\cite{FS,Zumino},
\bea
\bar Z^{ij}&\longrightarrow& \bar
Z'^{ij}=v^i_k v^j_l \bar Z^{kl}= \left(
\begin{array}{ccc}
0&-z&0\\
z&0&0\\
0&0&0
\end{array}
\right),
\label{e21}\\
Z_{ij}&\longrightarrow&
Z'_{ij}=\bar v_i^k \bar v_j^l Z_{kl}=
\left(
\begin{array}{ccc}
0&-\bar z& 0\\
\bar z&0&0\\
0&0&0
\end{array}
\right). \label{e22} \eea Correspondingly, (\ref{S2}) takes the
following form \be S_c=\int d\tau[
z(\theta_1^\alpha\dot\theta_{2\alpha}
-\theta_2^\alpha\dot\theta_{1\alpha}) +\bar
z(\bar\theta^1_{\dot\alpha}\dot{\bar\theta}{}^{2\dot\alpha}
-\bar\theta^2_{\dot\alpha}\dot{\bar\theta}{}^{1\dot\alpha})].
\label{e23}
\ee
Note that action (\ref{e23}) is nothing but the
Wess-Zumino term of the $\cN=2$ superparticle, which respects the
$SU(2)$ symmetry realized on the coordinates $\theta_{1\alpha}$,
$\theta_{2\alpha}$ and $\bar\theta^1_{\dot\alpha}$,
$\bar\theta^2_{\dot\alpha}$. There is also the $U(1)$ symmetry which
transforms the $\theta_{3\alpha}$ and $\bar \theta^3_{\dot\alpha}$
variables with a phase factor, \be \theta_{3\alpha} \to
e^{i\phi}\theta_{3\alpha},\qquad \bar \theta^3_{\dot\alpha}\to
e^{-i\phi}\bar \theta^3_{\dot\alpha}. \label{e24} \ee As a result,
we conclude that the Wess-Zumino term (\ref{S2}) breaks the internal
automorphisms symmetry $U(3)$ down to the $SU(2)\times U(1)$.

The Grassmann variables of the $\cN=3$ superspace can be rearranged
as
\be
\{ \theta_1^\alpha,\theta_2^\alpha,\theta_3^\alpha \}
\rightarrow \{\theta_\ui^\alpha,\theta_3^\alpha \},\qquad \{\bar
\theta^{1\dot\alpha},\bar \theta^{2\dot\alpha},\bar
\theta^{3\dot\alpha} \} \rightarrow
\{\bar\theta^{\ui\dot\alpha},\bar\theta^{3\dot\alpha} \},
\label{e161}
\ee
where the underlined indices $\ui,\uj$ are the
$SU(2)$ ones with the values $1,2$. Action (\ref{e23}) can now be
written as
\be
S_c=-\int
d\tau(z\varepsilon^{\ui\uj}\theta_\ui^\alpha\dot\theta_{\uj\alpha}
 -\bar z\varepsilon_{\ui\uj}\bar\theta^\ui_{\dot\alpha}\dot{\bar\theta}{}^{\uj\dot\alpha}),
\label{e167}
\ee
where $\varepsilon_{\ui\uj}$ is the antisymmetric
two-dimensional tensor, $\varepsilon_{12}=-\varepsilon^{12}=1$.
Action (\ref{e167}) has manifest $SU(2)$ invariance realized on the
indices $\ui,\uj$. Therefore a natural harmonic extension of such
superparticle model is given by the $SU(2)$ harmonic variables
$u^\pm_\ui$, \be u^\pm_\ui\in SU(2),\quad u^{\pm
\ui}=\varepsilon^{\ui\uj}u^\pm_\uj,\quad u^{+\ui}u^-_\ui=1.
\label{e163} \ee The Cartan forms on the $SU(2)$ group
\be
i\omega^{++}=u^+_\ui du^{+\ui},\quad
i\omega^{--}=du^-_\ui u^{-\ui},\quad
i\Theta=u^-_\ui du^{+\ui}
\label{e164}
\ee
are used to write down the particle action on a sphere
$S^2\sim SU(2)/U(1)$ \cite{Sorokin1,Sorokin2},
\be
S_{SU(2)}=2R^2\int d\tau\, e^{-1}\dot\omega^{++}\dot\omega^{--}
-\int d\tau\, \lambda(u^-_\ui u^{+\ui}-1) -\frac i2n\int
d\tau(u^-_\ui\dot u^{+\ui}-\dot u^-_\ui u^{+\ui}).
\label{e165}
\ee
Here $R$ is the radius of the sphere, $n$ is the electric charge of
the particle which couples to a magnetic field produced by a
monopole situated in the center of the sphere (see \cite{Sorokin1}
for details) and $\lambda$ is the Lagrange multiplier.

The harmonics $u^\pm_\ui$ convert $SU(2)$ indices $\ui,\uj,\ldots$
into $U(1)$ ones $\pm$, e.g., \be
\theta^\ui_{\alpha}\to\theta^\pm_\alpha=u^\pm_\ui\theta^\ui_\alpha,\qquad
\bar\theta^\ui_{\dot\alpha}\to
\bar\theta^\pm_{\dot\alpha}=u^\pm_\ui\bar\theta^\ui_{\dot\alpha}.
\label{t1} \ee Apart from usual complex conjugation, the harmonic
superspace
$\{x^m,\theta^\pm_\alpha,\bar\theta^\pm_{\dot\alpha},\theta_{3\alpha},
\bar\theta^3_{\dot\alpha},u^\pm_\ui \}$ has also
$\widetilde{\phantom{a}}$ conjugation defined as \footnote{This
conjugation is natural in the $\cN=2$ harmonic superspace with
coordinates
$\{x^m,\theta^\pm_\alpha,\bar\theta^\pm_{\dot\alpha},u^\pm_\ui\}$
\cite{Book} and acts on extra Grassmann variables
$\theta_{3\alpha}$, $\bar\theta^3_{\dot\alpha}$ as the usual complex
conjugation.} \bea &&\widetilde{u^\pm_\ui}=u^{\pm\ui},\qquad
\widetilde{u^{\pm\ui}}=-u^\pm_{\ui},\nonumber\\
&&\widetilde{\theta^\pm_\alpha}=\bar\theta^\pm_{\dot\alpha},\quad
\widetilde{\bar\theta^\pm_{\dot\alpha}}=-\theta^\pm_\alpha,\quad
\widetilde{\theta_{3\alpha}}=\bar\theta^3_{\dot\alpha},\quad
\widetilde{\bar\theta^3_{\dot\alpha}}=\theta_{3\alpha},
\label{conj2}
\eea
which leaves the action (\ref{e165}) invariant.

The action of a superparticle in harmonic
superspace $\{x^m,\theta^\pm_\alpha,\bar\theta^\pm_{\dot\alpha},\theta_{3\alpha},
\bar\theta^3_{\dot\alpha},u^\pm_\ui \}$ is given by the sum of
(\ref{S1}), (\ref{e167}) and (\ref{e165})
\be
S_{sp}+S_c+S_{SU(2)}=\int d\tau L_2,
\label{Sch}
\ee
where $L_2$ denotes the Lagrangian of the $\cN=3$ harmonic superparticle
with central charges.

\setcounter{equation}{0}
\section{Quantization of the $\cN=3$ superparticle without the central charge
term}
\subsection{Hamiltonian formulation and constraints}
We start with the action of the $\cN=3$ harmonic superparticle given
by (\ref{e29}). The corresponding Lagrangian reads
\bea
L_1&=&-\frac1{2e}\dot\omega^m\dot\omega_m+ \frac{R^2}{2 e}
 (\dot\omega^1_2\dot\omega^2_1+\dot\omega^2_3\dot\omega^3_2
 +\dot\omega^1_3\dot\omega^3_1)-\frac12e m^2\nn\\
&&-\frac{is_1}2(\dot u^1_i\bar u_1^i- u^1_i\dot{\bar u}{}_1^i)
+\frac{is_2}2(\dot u^3_i \bar u_3^i-u^3_i\dot{\bar
u}{}_3^i)\nn\\&&
+\sum_{I,J=1}^3 \lambda_I^J(u^I_i\bar u_J^i-\delta^I_J)
+ \Lambda(\varepsilon^{ijk}u^1_i u^2_j u^3_k
+ \varepsilon_{ijk}\bar u_1^i \bar u_2^j \bar u_3^k-2).
\label{L1}
\eea
The canonical momenta are defined by the Lagrangian (\ref{L1})
in a standard way,
\bea
p_m&=&-\frac{\partial L_1}{\partial\dot x^m}=e^{-1}\dot\omega^m,
\label{e34}\\
\pi^i_\alpha&=&\frac{\partial L_1}{\partial \dot\theta_i^\alpha}
 = ip_m(\sigma^m\bar\theta^i)_\alpha,
\label{e35}\\
\bar\pi_{i\dot\alpha}&=&\frac{\partial L_1}{\partial\dot{\bar\theta}{}^{i\dot\alpha}}
 = ip_m(\theta_i\sigma^m)_{\dot\alpha}=-(\pi^i_\alpha)^*,
\label{e36}\\
v^1_i&=&-\frac{\partial L_1}{\partial \dot{\bar u}{}_1^i}
 =\frac{R^2}{2e}(u^2_i\dot\omega^1_2+u^3_i\dot\omega^1_3)-\frac{is_1}2
 u^1_i,
\label{e37}\\
v^2_i&=&-\frac{\partial L_1}{\partial\dot{\bar u}{}_1^i}
=\frac{R^2}{2e}(u^1_i\dot\omega^2_1+u^3_i\dot\omega^2_3),
\label{e38}\\
v^3_i&=&-\frac{\partial L_1}{\partial\dot{\bar u}{}_3^i}
=\frac{R^2}{2e}(u^2_i\dot\omega^3_2+u^1_i\dot\omega^3_1)
+\frac{is_2}2 u^3_i,
\label{e39}\\
\bar v^i_1&=&-\frac{\partial L_1}{\partial\dot u{}^1_i}
=-\frac{R^2}{2e}(\bar u_2^i\dot\omega^2_1+\bar u_3^i\dot\omega^3_1)
+\frac{is_1}2  \bar u_1^i,
\label{e40}\\
\bar v^i_2&=&-\frac{\partial L_1}{\partial\dot u{}^2_i}
=-\frac{R^2}{2e}(\bar u_1^i\dot\omega^1_2+\bar u_3^i\dot\omega^3_2),
\label{e41}\\
\bar v^i_3&=&-\frac{\partial L_1}{\partial\dot u{}^3_i}
=-\frac{R^2}{2e}(\bar u_2^i\dot\omega^2_3+\bar u_1^i\dot\omega^1_3)
-\frac{is_2}2  \bar u_3^i.
\label{e42}
\eea
Note that equations
(\ref{e35},\ref{e36}) are the constraints since they do not allow to
express the Grassmann velocities through the corresponding momenta.

The standard mass-shell constraint appears from the equation of
motion for the einbein,
\be
0=\frac{\partial L_1}{\partial e}=\frac1{2e^2}[
\dot\omega^m\dot\omega_m-R^2
(\dot\omega^1_2\dot\omega^2_1+\dot\omega^2_3\dot\omega^3_2
+\dot\omega^1_3\dot\omega^3_1)-e^2m^2].
\label{mass-shell}
\ee

It is convenient to pass from the canonical harmonic momenta
$v^I_i$, $\bar v_I^i$ to the covariant ones $D^I_J$, $C^I_J$,
\be
D^I_J=u^I_i \bar v^i_J-\bar u^i_J v^I_i,\qquad C^I_J=u^I_i \bar
v^i_J+\bar u^i_J v^I_i, \label{e43} \ee which can be written
manifestly as \bea D^I_J&=&-R^2e^{-1}\dot\omega^I_J\qquad (I\ne J),
\label{e44}\\
D^1_1&=&is_1, \qquad D^2_2=0,\qquad D^3_3=-is_2,
\label{e45}\\
C^I_J&=&0 \qquad \forall I,J=1,2,3.
\label{e46}
\eea
Equations
(\ref{e45},\ref{e46}) are nothing but the constraints for the
harmonic variables. In terms of covariant momenta $D^I_J$, the
mass-shell constraint (\ref{mass-shell}) reads \be
p^mp_m-\frac1{R^2}(D^1_3D^3_1+D^1_2D^2_1+D^2_3D^3_2)-m^2=0.
\label{e47} \ee

Let us now define the Poisson brackets in a standard way:
\bea
&&[x^n,p_m]_P=-\delta^n_m,
\nn\label{e48}\\&&
\{\theta_i^\alpha,\pi^j_\beta
\}_P=-\delta_i^j\delta^\alpha_\beta,\qquad
\{\bar\theta^{i\dot\alpha},\bar\pi_{j\dot\beta}  \}_P=
 -\delta^i_j\delta^{\dot\alpha}_{\dot\beta},
\nn\label{e49}\\&&
[u^I_i,\bar v^j_J]_P=-\delta_i^j\delta^I_J,\qquad
[\bar u_I^i, v^J_j]_P=-\delta^i_j\delta_I^J.
\label{e50}
\eea
The harmonic covariant momenta (\ref{e44}) form $su(3)$ algebra
with the Cartan generators $S_1=D^1_1-D^2_2$,
$S_2=D^2_2-D^3_3$ under the Poisson brackets (\ref{e50}), e.g.,
\bea
&&[D^1_2,D^2_3]_P=D^1_3,\quad
[D^1_2,D^1_3]_P=0,\quad
[D^2_3,D^1_3]_P=0,\nn\\&&
[D^1_2,D^2_1]_P=S_1,\quad
[D^2_3,D^3_2]_P=S_1,\quad[S_1,D^1_2]_P=2D^1_2,\quad \mbox{etc}.
\label{e51}
\eea

There are the following non-trivial Poisson brackets between
the functions $C^I_J$, $D^2_2$ and the
constraints (\ref{h-constr1},\ref{h-constr3})
\be
[C^I_J,\chi^K_L]_P=2\delta^K_J\delta^I_L,\quad
[D^2_2,\chi_2]_P=2,\quad
[C^I_J,C^K_L]_P=\delta^K_J D^I_L-\delta^I_L D^K_J\equiv A^{IK}_{JL},
\label{e52}
\ee
which mean that they are second-class.
 Hence, they can be taken into account by introducing
the Dirac bracket \bea
[f,g\}_D&=&[f,g\}_P+\frac12[f,C^I_J]_P[\chi^J_I,g]_P
-\frac12[f,\chi^J_I]_P[C^I_J,g]_P \nn\\&&
+\frac12[f,D^2_2]_P[\chi_2,g]_P -\frac12[f,\chi_2]_P[D^2_2,g]_P
-\frac14[f,\chi^I_J]_P A^{JL}_{IK}[\chi^K_L,g]_P,
\label{e53}
\eea
where $f$ and $g$ are arbitrary phase space functions. The
applications of the Dirac and Poisson brackets resemble only the
harmonic variables and momenta, while for the other superspace
coordinates one can freely use the Poisson bracket instead of
(\ref{e53}).

Equations (\ref{e45}) contain the following first-class
constraints
\be
S_1-is_1\approx0,\qquad
S_2-is_2\approx0.
\label{e45_}
\ee

There are also the spinor constraints
\be D^i_\alpha=-\pi^i_\alpha
 +ip_m(\sigma^m\bar\theta^i)_\alpha\approx0,\qquad
\bar D_{i\dot\alpha} =\bar\pi_{i\dot\alpha}
-ip_m(\theta_i\sigma^m)_{\dot\alpha}\approx0, \label{e54} \ee which
anticommute non-trivially under the Poisson brackets (\ref{e50}),
\be \{D^i_\alpha,\bar D_{j\dot\alpha} \}_P=
-2i\delta^i_j\sigma^m_{\alpha\dot\alpha}p_m. \label{e55} \ee In
general, the constraints (\ref{e54}) are second-class. However, if
the dynamics of the superparticle is constrained to the surface
$p_mp^m\approx0$, the matrix in the rhs of (\ref{e55}) is degenerate
and the superparticle action is invariant under
$\kappa$-symmetry.\footnote{The $\kappa$-symmetry was first observed
in the model of a massive $\cN=2$, $d=4$ superparticle with the
central charge \cite{Luk1} and in the case of a massless $\cN=1$
superparticle in \cite{Siegel}. The group-theoretical and
geometrical origin of $\kappa$-symmetry as a manifestation of local
extended supersymmetry of the superparticle worldline was found in
\cite{stv}. This observation leads to the development of the
superembedding approach to the description of superbranes (see
\cite{superembedding} for a review and references).} In this case
both first-class and second-class constraints are entangled in
(\ref{e54}). As follows from (\ref{e47}), the condition
$p_mp^m\approx0$ is satisfied only for the massless superparticle,
$m=0$, with the following additional constraint for harmonic
variables
\be D^1_2D^2_1+D^2_3D^3_2+D^1_3D^3_1\approx0.
\label{add}
\ee
As we will show in the next section, this case corresponds
exactly to physical supermultiplets upon quantization.

Apart from the constraints considered above, it is meaningful to
introduce the following extra harmonic constraints
\cite{Sorokin1,Sorokin2}
\be D^1_2\approx0,\qquad
D^2_3\approx0,\qquad D^1_3\approx0.
\label{extra}
\ee
These
constraints are first-class and reduce the mass-shell condition
(\ref{e47}) to the physical one, \be p^mp_m-m^2\approx 0.
\label{extra1} \ee Equations (\ref{extra}) ``freeze'' the dynamics
of harmonic variables leaving only the motion of a particle in
$\{x^m, \theta_{i\alpha},\bar\theta^i_{\dot\alpha} \}$ superspace.
This fact emphasizes the unphysical meaning of the harmonic
variables in field theory. Indeed, upon quantization, these
constrains will yield the equations which eliminate an infinite
number of auxiliary fields with arbitrary number of $SU(3)$ indices
and leave only physical components.

The canonical Hamiltonian is defined using the momenta
(\ref{e34})--(\ref{e42}) via the Legendre transform, \be H_1=-\dot
x^m p_m -\dot u^I_i \bar v^i_I - \dot{\bar u}{}_I^i v^I_i
+\dot\theta_i^\alpha \pi^i_\alpha+\dot{\bar
\theta}{}^{i\dot\alpha}\bar\pi_{i\dot\alpha} -L_1, \label{e83} \ee
where $L_1$ is given by (\ref{L1}). Now we express the velocities
from (\ref{e34})--(\ref{e42}) and substitute them into the
Hamiltonian (\ref{e83}),
\be
H_1=-\frac e2(p^mp_m-m^2)+\frac
e{2R^2}(D^3_1D^1_3+D^2_1D^1_2+D^3_2D^2_3)
+\dot\omega^1_1(-S_1+is_1)+\dot\omega^3_3(S_2-is_2)-L_\lambda,
\label{e84}
\ee
where $L_\lambda$ corresponds to the last line in
the Lagrangian (\ref{L1}) with Lagrange multipliers. This Lagrangian
$L_\lambda$ is not essential when the harmonic constraints
(\ref{h-constr1},\ref{h-constr2}) are taken into account by the
Dirac bracket (\ref{e53}). Therefore we omit $L_\lambda$ further,
assuming the use of the Dirac bracket (\ref{e53}) in what follows.
Note also that the velocities $\dot\omega^1_1$, $\dot\omega^3_3$ can
not be eliminated from the Hamiltonian and remain arbitrary
functions. They play the role of Lagrange multipliers (as well as
the einbein $e$) and we denote them as $\mu(\tau)$, $\nu(\tau)$,
respectively. As a result we obtain the total Hamiltonian which is a
linear combination of first-class constraints, \be H_1=-\frac
e2[p^mp_m-\frac1{R^2}(D^3_1D^1_3+D^2_1D^1_2+D^3_2D^2_3)
-m^2]+\mu(-S_1+is_1)+\nu(S_2-is_2). \label{e85} \ee

The Hamiltonian equation of motion for any phase space coordinate
$f$ has the standard form \be \dot f=[f,H_1]_D. \label{e86} \ee Here
we do not write down these equations in detail, however they can be
easily figured out.

\subsection{Gupta-Bleuler quantization}
According to the postulates of canonical quantization, one replaces
the canonical momenta (\ref{e34})--(\ref{e42}) by the corresponding
differential operators, \be p_m\to i\frac\partial{\partial
x^m},\quad \pi^i_\alpha\to
-i\frac\partial{\partial\theta_i^\alpha},\quad \bar
\pi_{i\dot\alpha}\to-i\frac\partial{\partial\bar\theta^{i\dot\alpha}},
\quad v^I_i\to \frac\partial{\partial\bar u_I^i},\quad \bar
v_I^i\to\frac\partial{\partial u^I_i}.
\label{e100.2}
\ee
The spinor
constraints (\ref{e54}) turn into the covariant spinor
derivatives,\footnote{The operators $D^i_\alpha$, $\bar
D_{i\dot\alpha}$ are multiplied here also by $-i$ for convenience,
so that the operators (\ref{e101}) are related to each other by
complex conjugation rater than Hermitian one. The same concerns the
derivatives (\ref{a200})--(\ref{e205}) as well as the supercharges
(\ref{s-charges1},\ref{Q5}) and $U(1)$ charges $S_1$, $S_2$ in
(\ref{e109}).} \be D^i_\alpha=\frac\partial{\partial
\theta_i^\alpha} +i(\sigma^m\bar\theta^i)_\alpha\partial_m,\qquad
\bar D_{i\dot\alpha}=-\frac\partial{\partial
\bar\theta^{i\dot\alpha}}
-i(\theta_i\sigma^m)_{\dot\alpha}\partial_m, \label{e101} \ee and
the covariant harmonic momenta (\ref{e43}) lead to the covariant
harmonic derivatives, \be D^I_J=u^I_i\frac\partial{\partial u^J_i}-
\bar u_J^i\frac\partial{\partial \bar u_I^i}. \label{e102} \ee

Further we will use covariant spinor derivatives contracted with harmonics,
\be
D^I_\alpha=u^I_i D^i_\alpha,\qquad
\bar D_{I\dot\alpha}=\bar u_I^i \bar D_{i\dot\alpha},
\label{e104}
\ee
which satisfy the following algebra
\be
\{D^I_\alpha,\bar D_{J\dot\alpha} \}=-2i\delta^I_J\sigma^m_{\alpha\dot\alpha}
\partial_m,\quad
\{D^I_\alpha, D^J_\beta \}=0,\quad
\{\bar D_{I\dot\alpha},\bar D_{J\dot\beta} \}=0.
\label{e105}
\ee

The supercharges (\ref{supercharges}) are promoted to the operators
\be
Q^i_\alpha=-\frac\partial{\partial\theta_i^\alpha}
+i(\sigma^m\bar\theta^i)_\alpha\partial_m,\qquad \bar
Q_{i\dot\alpha}=\frac\partial{\partial\bar\theta^{i\dot\alpha}}
-i(\theta_i\sigma^m)_{\dot\alpha}\partial_m,
\label{s-charges1}
\ee
which form the $\cN=3$ superalgebra,
\be
\{Q^i_\alpha,\bar
Q_{j\dot\alpha} \}=2i\delta^i_j\sigma^m_{\alpha\dot\alpha}
\partial_m,\quad
\{Q^i_\alpha,Q^j_\beta \}=0,\quad\{\bar Q_{i\dot\alpha},\bar Q_{j\dot\beta} \}=0.
\label{N3-algebra}
\ee

The operators (\ref{e100.2}) should be realized in some Hilbert
space formed by the functions $|\Phi\rangle$, \be
|\Phi\rangle=\Phi(x^m,\theta_{i\alpha},\bar\theta^i_{\dot\alpha},
u). \label{e106} \ee Superfield (\ref{e106}) should satisfy some
equations of motion and constraints which originate from the
superparticle constraints. The superparticle has both first- and
second-class constraints. The first-class constraints form closed
algebra under the Poisson or Dirac bracket. Therefore, they all
should be imposed on states (\ref{e106}),
\bea
&&S_1\Phi^{(s_1,s_2)}=s_1 \Phi^{(s_1,s_2)},\qquad
S_2\Phi^{(s_1,s_2)}=s_2 \Phi^{(s_1,s_2)},
\label{e109}\\
&&[\partial^m\partial_m +\frac1{R^2}X+m^2]\Phi^{(s_1,s_2)}=0,
\label{e110}
\eea
where \footnote{Here we use a particular ordering
of the operators $D^I_J$ although other orderings are also
possible.} \be X=D^2_1D^1_2+D^3_2 D^2_3+ D^3_1 D^1_3. \label{e111}
\ee Equations (\ref{e109}) mean that the superfield
$\Phi^{(s_1,s_2)}$ is a function of harmonic variables with definite
$U(1)$ charges. Note that the equations (\ref{e109}) covariantly
constrain the $SU(3)$ harmonic dynamics to the one on a coset
$SU(3)/(U(1)\times U(1))$. Equation (\ref{e110}) is the mass-shell
constraint which gives Klein-Gordon-like equations for all physical
fields. Note that the zero modes of the operator (\ref{e111})
(states which are annihilated by this operator,
$X\Phi_0^{(s_1,s_2)}=0$) satisfy the standard Klein-Gordon equation
without a harmonic term,
\be
(\partial^m\partial_m+m^2)\Phi^{(s_1,s_2)}_0=0. \label{e112} \ee

The second-class constraints are accounted either by constructing
the corresponding Dirac bracket or by applying the Gupta-Bleuler
method. In our case, the second-class harmonic constraints
(\ref{e52}) are taken into account by the Dirac bracket (\ref{e53}),
while the spinorial ones (\ref{e54}) should be considered \`a la
Gupta-Bleuler. It means that they have to be divided into two
complex conjugate subsets with weakly commutative constraints in
each subset. There are four different ways of separation of
derivatives (\ref{e101}) or (\ref{e104}) into such subsets:
\begin{subequations}
\begin{eqnarray}
\{D^1_\alpha,D^2_\alpha, D^3_\alpha \}&\cup&
 \{\bar D_{1\dot\alpha},\bar D_{2\dot\alpha},\bar D_{3\dot\alpha}
 \},
\label{i}\\
\{D^1_\alpha,\ \bar D_{2\dot\alpha},\ \bar D_{3\dot\alpha}
\}&\cup&
\{\bar D_{1\dot\alpha},\ D^2_\alpha,\ D^3_\alpha \},
\label{ii}\\
\{D^1_\alpha,\ D^2_\alpha,\ \bar D_{3\dot\alpha}
\}&\cup&
\{\bar D_{1\dot\alpha},\ \bar D_{2\dot\alpha},\ D^3_\alpha \},
\label{iii}\\
\{\bar D_{1\dot\alpha},\ D^2_\alpha,\ \bar D_{3\dot\alpha} \}
&\cup&
\{ D^1_\alpha,\ \bar D_{2\dot\alpha},\ D^3_\alpha \}.
\label{iv}
\end{eqnarray}
\end{subequations}
Different choices of subsets (\ref{i})--(\ref{iv}) lead to
different types of quantization of the superparticle. In the following
subsections we consider them separately.

\subsubsection{$\cN=3$ gravitino multiplet}
In this subsection we will show that the separation of fermionic
constraints (\ref{i}) leads to the $\cN=3$ gravitino multiplet with
the highest helicity 3/2. First, we consider the massive case,
$m\ne0$, where the spinor constraints $D^i_\alpha\approx0$, $\bar
D_{i\dot\alpha}\approx0$ are second-class since the matrix in the
rhs of (\ref{e55}) is invertible. There is no $\kappa$-symmetry in
the model and, hence, no extra constraints. According to (\ref{i}),
the physical state is annihilated only by the derivative $\bar
D_{i\dot\alpha}$, while $D^i_\alpha$ kills the conjugate superfield,
\be
\bar D_{i\dot\alpha} \Phi^{(s_1,s_2)}=0,\qquad D^i_\alpha\bar
\Phi^{(s_1,s_2)}=0.
\label{e113}
\ee
The dynamics of such a field is
described by the set of equations (\ref{e109},\ref{e110},\ref{e113})
which take into account all the superparticle constraints. Note that
equations (\ref{e113}) are nothing but the chirality conditions for
the field $\Phi^{(s_1,s_2)}$. Therefore we refer to such a
quantization as a chiral quantization.

The non-zero modes of the operator $X$ propagate analogously to zero
ones. It means that the superfield $\Phi^{(s_1,s_2)}$ describes an
unphysical multiplet with infinite number of component fields. To
make it physical, we impose the additional harmonic constraints
(\ref{extra}),
\be D^1_2 \Phi^{(s_1,s_2)}=0,\quad
D^2_3\Phi^{(s_1,s_2)}=0,\quad D^1_3 \Phi^{(s_1,s_2)}=0.
\label{e114}
\ee
In general, function $\Phi^{(s_1,s_2)}$ is given by a series in
harmonic variables. Equations (\ref{e114}) reduce this series to a
monomial \be \Phi^{(s_1,s_2)}= \bar u_3^{i_1}\ldots\bar
u_3^{i_{s_2}}u^1_{j_1}\ldots u^1_{j_{s_1-s_2}} \Phi_{i_1\ldots
i_{s_2}}^{j_1\ldots j_{s_1-s_2}}, \label{e125} \ee where
$\Phi_{i_1\ldots i_{s_2}}^{j_1\ldots j_{s_1-s_2}}$ is a totally
symmetric traceless tensor. Indeed, $D^1_2$, $D^2_3$, $D^1_3$ are
raising operators in the $su(3)$ algebra which define the highest
weight vector (\ref{e125}) \cite{Delduc,Ferrara}. As a result, we
obtain the chiral superfield with fixed number of $SU(3)$ indices
(symmetric and traceless) on mass-shell, \be \bar
D_{i\dot\alpha}\Phi_{i_1\ldots i_{s_2}}^{j_1\ldots
j_{s_1-s_2}}=0,\qquad (\partial^m\partial_m+m^2)\Phi_{i_1\ldots
i_{s_2}}^{j_1\ldots j_{s_1-s_2}}=0. \label{e115} \ee

Let us consider, e.g., the simplest representation $\Phi$ without
$SU(3)$ indices that corresponds to the choice of $U(1)$ charges
$s_1=s_2=0$. The solution of the chirality condition
$ \bar D_{i\dot\alpha}\Phi=0$ is most naturally given in chiral coordinates
$y^m=x^m+i\theta_i\sigma^m\bar\theta^i$,
\bea
\Phi(y,\theta)&=&\phi+\theta_i^\alpha\psi^i_\alpha
+\theta_i^\alpha\theta_j^\beta\varepsilon^{ijk}F_{k\,(\alpha\beta)}
+\theta^\alpha_i\theta_{j\alpha} d^{(ij)}\nn\\&&
+\theta_i^\alpha\theta_{j\alpha}\theta_k^\beta\varepsilon^{ikl}
 \chi^j_{l\,\beta}
+\theta_i^\alpha\theta_j^\beta\theta_k^\gamma \varepsilon^{ijk}
 T_{(\alpha\beta\gamma)}\nn\\&&
+\theta_i^\alpha\theta_{j\alpha}\theta_k^\beta\theta_{l\beta}
 \varepsilon^{ikr}\varepsilon^{jlm}S_{(rm)}
+\theta_i^\alpha\theta_j^\beta\theta_k^\gamma\theta_{l\gamma}
 \varepsilon^{ijk}G^l_{(\alpha\beta)}\nn\\&&
+\theta^\alpha_i\theta^\beta_j\theta^\gamma_k
 \theta_{l\alpha}\theta_{m\beta}\varepsilon^{ijk}\varepsilon^{lmn}
  \rho_n^\gamma
+\theta_i^\alpha\theta_j^\beta\theta_k^\gamma
 \theta_{l\alpha}\theta_{m\beta}\theta_{n\gamma}
 \varepsilon^{ijk}\varepsilon^{lmn} U,
\label{e115.4}
\eea
where all components depend on $y^m$. Note that
both bosonic and fermionic components in (\ref{e115.4}) satisfy the
Klein-Gordon equation owing to (\ref{e115}), but there are no Dirac
equations for spinors. Therefore, such a multiplet is unphysical. We
assume that the mass should be introduced not directly but through a
central charge, as is shown in the next section. Therefore for the
rest of this section we consider only the massless case, $m=0$.

In the massless case the superparticle action (\ref{S1}) is well
known to respect the $\kappa$-symmetry since the matrix in the rhs
of (\ref{e55}) is degenerate. Half of the constraints
$D^i_\alpha\approx 0$, $\bar D_{i\dot\alpha}\approx0$ turn into
first-class ones. But if we deal with the harmonic superparticle
with the action (\ref{e25}), the standard mass-shell constraint
(\ref{e112}) is replaced by the equation (\ref{e110}) with the
harmonic contribution due to operator $X$. Therefore, for the states
out from the kernel of operator $X$, the matrix
$\sigma^m_{\alpha\dot\alpha}\partial_m$ is invertible and the
constraints $D^i_\alpha\approx 0$, $\bar D_{i\dot\alpha}\approx0$
still belong to the second class. As explained before, the
$\kappa$-symmetry of harmonic superparticle is restored and one half
of these second-class constraints turn into first-class ones, if the
dynamics is constrained by (\ref{add}). Upon quantization,
constraint (\ref{add}) is imposed on the states implying the
condition $X\,\Phi=0$. Namely such states are interesting from the
physical point of view.

In particular, on the surface of constraints (\ref{extra}), the
condition (\ref{add}) is satisfied and the kinetic part of the
action for harmonics (\ref{e26}) can be omitted. Therefore the
dynamics is described effectively by the action
\be
S=-\frac12\int
d\tau\frac{\dot\omega^m\dot\omega_m}{e}+S_{WZ}+S_\lambda,
\label{e116} \ee which is invariant under the following
transformations of $\kappa$-symmetry \bea
\delta_\kappa\theta_{i\alpha}&=&
 -ip_m(\sigma^m\bar\kappa_i)_\alpha,\qquad
\delta_\kappa\bar\theta^i_{\dot\alpha}=
 ip_m(\kappa^i\sigma^m)_{\dot\alpha},
\label{e117}\nn\\
\delta_\kappa x^m&=&i\delta_\kappa\theta_i\sigma^m\bar\theta^i
 -i\theta_i\sigma^m \delta_\kappa\bar\theta^i,
\label{e118}\nn\\
\delta_\kappa e&=&-4(\bar\kappa_i\dot{\bar\theta}^i+
\dot\theta_i\kappa^i),
\label{e119}
\eea
where
$\kappa^i_\alpha(\tau)$, $\bar\kappa_{i\dot\alpha}(\tau)$ are
anticommuting local parameters. Note that the harmonic terms
$S_{WZ}$, $S_\lambda$, given by (\ref{e27}) and (\ref{e28})
respectively, do not violate the $\kappa$-symmetry as the harmonics
do not transform, $\delta_\kappa u^I_i=0$, $\delta_\kappa \bar
u_I^i=0$. Therefore the fields $\Phi^{(s_1,s_2)}$ with all values of
$U(1)$ charges should obey the constraints originating from the
$\kappa$-symmetry.

The transformations (\ref{e119}) are generated by the Poisson
brackets of coordinates with the following first-class
constraints,
\be
\psi_{i\alpha}=ip_m\sigma^m_{\alpha\dot\alpha}\bar
D_i^{\dot\alpha}\approx0,\qquad
\bar\psi^i_{\dot\alpha}=-ip_m\sigma^m_{\alpha\dot\alpha}D^{i\alpha}\approx0.
\label{e120}
\ee
Upon quantization, (\ref{e120}) turn into the differential operators,
\be
\psi_{i\alpha}=-\partial_m\sigma^m_{\alpha\dot\alpha}\bar
D_i^{\dot\alpha}, \qquad
\bar\psi^i_{\dot\alpha}=\partial_m\sigma^m_{\alpha\dot\alpha}
D^{i\alpha},
\label{e121}
\ee
where $D^i_\alpha$, $\bar D_{i\dot\alpha}$ are the covariant
spinor derivatives (\ref{e101}). These differential operators
should annihilate the physical states,
\be
\partial_m\sigma^m_{\alpha\dot\alpha}
D^{i\alpha} \Phi^{(s_1,s_2)}=0.
\label{e122}
\ee
Recall that the
superfield $\Phi^{(s_1,s_2)}$ is a chiral $\cN=3$ superfield
(\ref{e113}) constrained by (\ref{e109},\ref{e112},\ref{e114}).
First-class constraints (\ref{e112}) and (\ref{e122}) arise from the
following one as the integrability conditions (cf.
\cite{Sorokin1,Sorokin2} in $\cN=2$ case) \be
D^{i\alpha}D^j_{\alpha}\Phi^{(s_1,s_2)}=0. \label{e123} \ee It is
the constraint (\ref{e123}) which leads to the correct component
structure of the multiplet and eliminates all auxiliary fields in
the decomposition (\ref{e115.4}), despite the fact that it is
stronger than (\ref{e112}) and (\ref{e122}). Therefore, we use
further (\ref{e123}) rather than
(\ref{e112},\ref{e122}).\footnote{In fact, (\ref{e123}) is a
consequence of $\kappa$-symmetry constraints (\ref{e120}) since on
the surface of constraints (\ref{e54},\ref{extra1}) the following
relations $\bar d_{ij}\equiv\theta_i^\alpha\psi_{j\alpha}\approx
-\bar D_{i\dot\alpha}\bar D_j^{\dot\alpha}$,
$d^{ij}\equiv\bar\theta^i_{\dot\alpha}\bar\psi^{j\dot\alpha}\approx
-D^{i\alpha}D^j_\alpha$ hold. It is easy to see that the constraints
$\bar d_{ij}\approx0$, $d^{ij}\approx0$ are the generators
$\kappa$-transformations (\ref{e119}) with the parameters
$\kappa^i_\alpha=k^{ij}\theta_{j\alpha}$,
$\bar\kappa_{i\dot\alpha}=\bar k_{ij}\bar\theta^j_{\dot\alpha}$,
where $k^{ij}$, $\bar k_{ij}$ are new bosonic local parameters.
Analogously, the constraint (\ref{e230}) follows from
$\kappa$-symmetry ones (\ref{e227}).}

Let us summarize all the equations for the superfield
$\Phi^{(s_1,s_2)}$ in a single list,
\be
\begin{array}l
S_1\Phi^{(s_1,s_2)}=s_1 \Phi^{(s_1,s_2)},\qquad
S_2\Phi^{(s_1,s_2)}=s_2 \Phi^{(s_1,s_2)},\\
D^1_2 \Phi^{(s_1,s_2)}=
D^2_3\Phi^{(s_1,s_2)}=
D^1_3 \Phi^{(s_1,s_2)}=0,\\
\bar D_{i\dot\alpha} \Phi^{(s_1,s_2)}=0,\\
D^{i\alpha}D^j_{\alpha}\Phi^{(s_1,s_2)}=0.
\end{array}
\label{e124}
\ee
The solution of the pure harmonic constraints in the first two
lines of (\ref{e124}) is given by (\ref{e125}).
The other constraints in (\ref{e124}) give
the chirality and linearity conditions,
\be
\bar D_{k\dot\alpha}\Phi_{i_1\ldots i_{s_2}}^{j_1\ldots
j_{s_1-s_2}}=0,\qquad
D^{k\alpha}D^l_\alpha \Phi_{i_1\ldots i_{s_2}}^{j_1\ldots
j_{s_1-s_2}}=0.
\label{e126}
\ee

Consider the equations (\ref{e126}) in the
simplest case of the scalar superfield $\Phi$ without $SU(3)$
indices,
\be
\bar D_{i\dot\alpha}\Phi=0,\qquad
D^{i\alpha}D^j_\alpha \Phi=0.
\label{e127}
\ee
The component structure of
a general chiral superfield $\Phi$ is given by
(\ref{e115.4}). The linearity condition eliminates all unphysical
components,
\be
\Phi=\phi+\theta_i^\alpha\psi^i_\alpha
+\theta_i^\alpha\theta_j^\beta\varepsilon^{ijk}F_{k(\alpha\beta)}
+\theta_i^\alpha\theta_j^\beta\theta_k^\gamma
 \varepsilon^{ijk}T_{(\alpha\beta\gamma)}.
\label{e128}
\ee
Owing to (\ref{e127}), the component fields in (\ref{e128}) satisfy the standard d'Alembert,
Weyl and Maxwell equations,
\bea
\square\phi=0&\phantom{aa}&\mbox{1 complex scalar},\nn\\
\sigma^{m}{}^\alpha{}_{\dot\alpha}\partial_m\psi^i_\alpha=0&&
\mbox{3 Weyl spinors},\nn\\
\sigma^{m}{}^\alpha{}_{\dot\alpha}\partial_m F_{k(\alpha\beta)}=0 && \mbox{3 vectors},\nn\\
\sigma^{m}{}^\alpha{}_{\dot\alpha}\partial_m
T_{(\alpha\beta\gamma)}=0 &&\mbox{1 gravitino.}
\label{e129}
\eea
Therefore, superfield $\Phi$ describes the $\cN=3$ supersymmetric
gravitino multiplet.

\subsubsection{$\cN=3$ supergauge multiplet}
In this subsection we will show that the separations of constraints
(\ref{ii})--(\ref{iv}) lead to the $\cN=3$ supergauge multiplet
\cite{Sorokin1,Ferrara}. First, we will analyze the separation
(\ref{ii}) in details and then will comment on the (\ref{iii}) and
(\ref{iv}) cases.

Recall that superfield $\Phi^{(s_1,s_2)}$ satisfies the first-class
constraints (\ref{e109},\ref{e110}). Now we impose also the
spinorial constraints from the first subset in (\ref{ii}), \be
D^1_\alpha \Phi^{(s_1,s_2)}=0,\quad \bar D_{2\dot\alpha}
\Phi^{(s_1,s_2)}=0,\quad \bar D_{3\dot\alpha}\Phi^{(s_1,s_2)}=0.
\label{e133}
\ee
Constraints (\ref{e133}) show that superfield
$\Phi^{(s_1,s_2)}$ is analytic, i.e., it is short in the component
expansion. Therefore, we refer to such type of quantization as an
analytic quantization. Note that different types of analytic
subspaces in full $\cN=3$ harmonic superspace introduced in
\cite{Ferrara} correspond to different subsets of Grassmann
derivatives (\ref{ii})--(\ref{iv}) annihilating the state.

It is easy to observe that the spinor derivatives in
(\ref{e133}) do not commute with the operator $X$ given by (\ref{e111}),
\be
[D^1_\alpha,X]=-D^2_\alpha D^1_2-D^3_\alpha D^1_3,\quad
[\bar D_{2\dot\alpha},X]=\bar D_{1\dot\alpha}D^1_2
 +D^3_2 \bar D_{3\dot\alpha},\quad
[\bar D_{3\dot\alpha},X]=\bar D_{2\dot\alpha}D^2_3+\bar
D_{1\dot\alpha}D^1_3.
\label{e134}
\ee
Therefore, the analytic
quantization is consistent only if the state $\Phi^{(s_1,s_2)}$
satisfies extra harmonic constraints (\ref{e114}). Owing to these
constraints, the operators in the rhs of (\ref{e134}) vanish on the
state while the constraint (\ref{e110}) has no harmonic part and
turns into the usual mass-shell constraint ($m=0$), \be
\square\Phi^{(s_1,s_2)}=0. \label{e135} \ee Equations (\ref{e114})
leave only zero modes of the operator $X$, which are massless. For
such modes, the harmonic variables are not dynamical and the action
(\ref{e116}) possesses $\kappa$-symmetry (\ref{e119}). Let us
project the generators of $\kappa$-symmetry (\ref{e121}) with
harmonics, \be
\psi_{I\alpha}=-\partial_m\sigma^m_{\alpha\dot\alpha}\bar
D_I^{\dot\alpha},\qquad
\bar\psi^I_{\dot\alpha}=\partial_m\sigma^m_{\alpha\dot\alpha}D^{I\alpha}.
\label{e136} \ee Operators (\ref{e136}) should annihilate the state
$\Phi^{(s_1,s_2)}$ since the constraints (\ref{e120}) are
first-class. Owing to the analyticity (\ref{e133}), it is sufficient
to impose three of the six operators (\ref{e136}) as the
constraints, \be
\partial_m\sigma^m_{\alpha\dot\alpha}\bar D_1^{\dot\alpha}
\Phi^{(s_1,s_2)}=0,\quad
\partial_m\sigma^m_{\alpha\dot\alpha}D^2_\alpha
\Phi^{(s_1,s_2)}=0,\quad
\partial_m\sigma^m_{\alpha\dot\alpha}D^3_\alpha
\Phi^{(s_1,s_2)}=0. \label{e137} \ee These constraints (\ref{e137}),
as well as the mass-shell condition (\ref{e135}), follow from the
more general ones \be (\bar D_1)^2\Phi^{(s_1,s_2)}=0,\quad
(D^2)^2\Phi^{(s_1,s_2)}=0,\quad (D^3D^2)\Phi^{(s_1,s_2)}=0,\quad
(D^3)^2\Phi^{(s_1,s_2)}=0.
\label{e138}
\ee
In spite of constraints
(\ref{e138}) being stronger than (\ref{e135}) and (\ref{e137}), they
should be also imposed on the state by the same reasons as
constraint (\ref{e123}), obtained for the $\cN=3$ gravitino
multiplet.

We summarize all the constraints for the superfield
$\Phi^{(s_1,s_2)}$ in a single list,
\be
\begin{array}l
S_1\Phi^{(s_1,s_2)}=s_1 \Phi^{(s_1,s_2)},\qquad
S_2\Phi^{(s_1,s_2)}=s_2 \Phi^{(s_1,s_2)},\\
D^1_2 \Phi^{(s_1,s_2)}=
D^2_3\Phi^{(s_1,s_2)}=
D^1_3 \Phi^{(s_1,s_2)}=0,\\
D^1_\alpha \Phi^{(s_1,s_2)}=
\bar D_{2\dot\alpha} \Phi^{(s_1,s_2)}=
\bar D_{3\dot\alpha}\Phi^{(s_1,s_2)}=0,\\
(\bar D_1)^2\Phi^{(s_1,s_2)}=
(D^2)^2\Phi^{(s_1,s_2)}=
(D^3D^2)\Phi^{(s_1,s_2)}=
(D^3)^2\Phi^{(s_1,s_2)}=0.
\end{array}
\label{e139}
\ee
Further we consider some examples of solutions of these constraints
for the lowest values of $U(1)$ charges.

Let $s_1=s_2=0$. The corresponding state is described by the
chargeless superfield $\Phi$. It is easy to show that under
constraints (\ref{e139}), this superfield is just a constant,
$\Phi=const$. Therefore this case is trivial.

The next case with $s_1=1$, $s_2=0$ was considered in
\cite{Sorokin1}. We denote the corresponding superfield
by $\Phi^{(1,0)}= W^1$. As a consequence of  (\ref{e139}),
it satisfies the following equations of motion and constraints

\be
\begin{array}l
D^1_2 W^1=D^2_3 W^1=D^1_3 W^1=0,\\
D^1_\alpha W^1=\bar D_{2\dot\alpha} W^1=
 \bar D_{3\dot\alpha} W^1=0,\\
(\bar D_1)^2 W^1=(D^2)^2 W^1=(D^3)^2 W^1
=(D^2D^3) W^1=0.
\end{array}
\label{e141}
\ee
Equations (\ref{e141}) are known to describe the
$\cN=3$ superfield strength of the gauge multiplet
\cite{Sorokin1,Ferrara}. The component structure of $W^1$ can be
most easily found in the analytic coordinates, \be y_A^m=x^m-
 i(\theta_1^\alpha\sigma^m\bar\theta^{1\dot\alpha}-
 \theta_3^\alpha\sigma^m\bar\theta^{3\dot\alpha}
 -\theta_2^\alpha\sigma^m\bar\theta^{2\dot\alpha}),
\label{e143}
\ee
in which the spinor derivatives $D^1_\alpha$, $\bar
D_{2\dot\alpha}$, $\bar D_{3\dot\alpha}$ take the most simple form,
\be D^1_\alpha=\frac\partial{\partial\theta_1^\alpha},\quad \bar
D_{2\dot\alpha}=-\frac\partial{\partial\bar\theta^{2\dot\alpha}},
\quad \bar
D_{3\dot\alpha}=-\frac\partial{\partial\bar\theta^{3\dot\alpha}}.
\label{e144} \ee Then we have, \be
W^1=\phi^1+\bar\theta^{1\dot\alpha}\bar\lambda_{\dot\alpha}
+\theta_2^\alpha\lambda_{3\alpha} -\theta_3^\alpha\lambda_{2\alpha}
-i\theta_2^\alpha\bar\theta^{1\dot\alpha}\sigma^m_{\alpha\dot\alpha}
\partial_m \phi^2
+\theta_2^\alpha\theta_3^\beta F_{(\alpha\beta)}
-i\bar\theta^{1\dot\alpha}\theta_2^\alpha\theta_3^\alpha
\sigma^m_{(\alpha\dot\alpha}\partial_m\lambda_{1\beta)},
\label{e145}
\ee
where $\phi^I=u^I_i\phi^i$ is a triplet of complex
scalars, $\lambda_{I\alpha}=\bar u_I^i\lambda_{i\alpha}$ is a
triplet of Weyl spinors, $\bar \lambda_{\dot\alpha}$ is also a Weyl
spinor, and $F_{(\alpha\beta)}$ is a Maxwell field strength. All
these components satisfy the corresponding free equations of motion.

Let us consider briefly the separation of constraints (\ref{iii})
leading to other superfield realizations. Consideration for the case
when the superfield $\Phi^{(s_1,s_2)}$ satisfies the following
Grassmann shortness conditions \be D^1_\alpha
\Phi^{(s_1,s_2)}=0,\quad D^2_\alpha \Phi^{(s_1,s_2)}=0,\quad \bar
D_{3\dot\alpha} \Phi^{(s_1,s_2)}=0
\label{e146}
\ee
can be done
similarly as in the previous case. The physically interesting
representation appears if the $U(1)$ charges take the values
$s_1=0$, $s_2=1$. We denote such a superfield by $\Phi^{(0,1)}=\bar
W_3$. It has the following equations of motion and constraints \be
\begin{array}l
D^1_2 \bar W_3=D^2_3\bar W_3=D^1_3\bar W_3=0,\\
D^1_\alpha \bar W_3 = D^2_\alpha \bar W_3=\bar D_{3\dot\alpha} \bar
W_3=0,\\
(\bar D_1)^2\bar W_3=(\bar D_2)^2 \bar W_3=(\bar D_1\bar D_2)\bar W_3
=(D^3)^2 \bar W_3=0,
\end{array}
\label{e148}
\ee
which describe the $\cN=3$ Maxwell multiplet as
well \cite{Sorokin1,Ferrara}. In particular, in the coordinates
$({y'}_A^m,\bar\theta^1, \bar\theta^2,\theta_3)$, \be {y'}_A^m=x^m-
 i(\theta_1^\alpha\sigma^m\bar\theta^{1\dot\alpha}-
 \theta_3^\alpha\sigma^m\bar\theta^{3\dot\alpha}
 +\theta_2^\alpha\sigma^m\bar\theta^{2\dot\alpha}),
\label{e149}
\ee
${\bar W_3}$ has the following component field
decomposition \bea \bar
W_3&=&\bar\phi_3-\theta_3^\alpha\lambda_\alpha
+\bar\theta^1_{\dot\alpha}\bar\lambda^{2\dot\alpha}
-\bar\theta^2_{\dot\alpha}\bar\lambda^{1\dot\alpha}
+i\theta_3^\alpha\bar\theta^{1\dot\alpha}
 \sigma^m_{\alpha\dot\alpha}\partial_m\bar\phi_1
-i\theta_3^\alpha\bar\theta^{2\dot\alpha}
 \sigma^m_{\alpha\dot\alpha}\partial_m\bar\phi_2\nn\\&&
-\bar\theta^{1\dot\alpha}\bar\theta^{2\dot\beta}
 \bar F_{(\dot\alpha\dot\beta)}
+i\bar\theta^1_{\dot\alpha}\bar\theta^2_{\dot\beta}
 \theta_3^\alpha\sigma^m{}_\alpha{}^{(\dot\alpha}
 \partial_m\bar\lambda^{3\dot\beta)}.
\label{e150}
\eea
Here, $\bar\phi_3=\bar u_3^i\bar\phi_i$ is a
triplet of complex scalars, $\lambda_\alpha$ is a Weyl spinor,
$\bar\lambda^I_{\dot\alpha}$ is a triplet of Weil spinors, and $\bar
F_{(\dot\alpha\dot\beta)}$ is a Maxwell field strength. It is easy
to see that superfields (\ref{e145}) and (\ref{e150}) are related to
each other by the conjugation (\ref{e17}).

In concluding this subsection, let us briefly comment on the last
case of constraints (\ref{iv}) on the example of a superfield
$\Phi^{(-1,1)}\equiv W^2$, \be D^2_\alpha W^2=0,\quad \bar
D_{1\dot\alpha}W^2=0,\quad \bar D_{3\dot\alpha}W^2=0. \label{e151}
\ee It is easy to see that the harmonic derivatives $D^1_3$,
$D^2_1$, $D^2_3$ commute with $D^2_\alpha$, $\bar D_{1\dot\alpha}$,
$\bar D_{3\dot\alpha}$ and therefore should also annihilate this
superfield, \be D^1_3 W^2=0,\quad D^2_1 W^2=0,\quad D^2_3 W^2=0.
\label{e152} \ee The $\kappa$-symmetry leads to the following
linearity constraints \be (D^1)^2W^2=(D^1 D^3)W^2=(D^3)^2 W^2=(\bar
D_2)^2W^2=0. \label{e153} \ee Such a superfield under  constraints
(\ref{e151})--(\ref{e153}) also describes the $\cN=3$ Maxwell
multiplet which is equivalent to (\ref{e145}). The component
structure of $W^2$ is similar to (\ref{e145}) with the change of
index from $1$ to $2$.

\setcounter{equation}{0}
\section{Quantization of the $\cN=3$ superparticle with a central charge
term}
\subsection{Hamiltonian formulation and constraints}
Let us consider the $\cN=3$ harmonic superparticle with central
charges described by the action (\ref{Sch}). The Lagrangian $L_2$
reads \bea L_2&=&-\frac1{2e}\dot\omega^m\dot\omega_m-\frac12em^2
-(z\varepsilon^{\ui\uj}\theta_\ui\dot\theta_\uj
 -\bar z\varepsilon_{\ui\uj}\bar\theta^\ui\dot{\bar\theta}{}^\uj)
 \nn\\&&
+2R^2 e^{-1}\dot\omega^{++}\dot\omega^{--} - \lambda(u^-_\ui
u^{+\ui}-1) -\frac i2n(u^-_\ui\dot u^{+\ui}-\dot u^-_\ui u^{+\ui}).
\label{L2}
\eea
Recall that the underlined indices $\ui,\uj,\ldots$
denote $SU(2)$ ones with the values $1,2$. The consequent
quantization of this model is similar to the one for the $\cN=2$
superparticle in harmonic superspace \cite{Sorokin1,Sorokin2}.
Therefore we follow the same steps keeping, however, basic details
of calculations.

The Lagrangian (\ref{L2}) defines the following canonical momenta
for superspace coordinates
\bea
p_m&=&-\frac{\partial L_2}{\partial \dot x^m}=e^{-1}\dot\omega_m,
\label{e170}\\
\pi^\ui_\alpha&=&\frac{\partial L_2}{\partial\dot\theta_\ui^\alpha}
=ip_m(\sigma^m\bar\theta^\ui)_\alpha+z\varepsilon^{\ui\uj}\theta_{\uj\alpha},
\label{e171}\\
\bar \pi_{\ui\dot\alpha}&=&\frac{\partial L_2}{\partial\dot{\bar\theta}{}^{\ui\dot\alpha}}
=ip_m(\theta_\ui\sigma^m)_{\dot\alpha}+\bar z\varepsilon_{\ui\uj}
 \bar\theta^\uj_{\dot\alpha},
\label{e172}\\
\pi^3_\alpha&=&\frac{\partial L_2}{\partial\dot\theta_3^\alpha}
=ip_m(\sigma^m\bar\theta^3)_\alpha,
\label{e174}\\
\bar \pi_{3\dot\alpha}&=&\frac{\partial L_2}{\partial\dot{\bar\theta}{}^{3\dot\alpha}}
=ip_m(\theta_3\sigma^m)_{\dot\alpha},
\label{e175}\\
v^{+\ui}&=&-\frac{\partial L_2}{\partial \dot u_\ui^-}
=2R^2e^{-1}u^{-\ui}i\dot\omega^{++}-\frac12inu^+_\ui,
\label{e176}\\
v^-_\ui&=&-\frac{\partial L_2}{\partial\dot u^{+\ui}}
=2R^2e^{-1}u^+_\ui i\dot\omega^{--}+\frac12inu^-_\ui.
\eea
Following \cite{Sorokin1,Sorokin2}, we introduce the covariant harmonic
momenta
\bea
D^{++}&=&u^+_\ui v^{+\ui}=-2iR^2e^{-1}\dot\omega^{++},
\qquad
D^{--}=v^-_\ui u^{-\ui}=-2iR^2e^{-1}\dot\omega^{--},
\label{e177}\label{e178}\\
D^0&=&v^-_\ui u^{+\ui}-u^-_\ui v^{+\ui}=in,
\qquad
\chi_2=v^-_\ui u^{+\ui}+u^-_\ui v^{+\ui}=0
\label{e180}
\eea
and define the Poisson brackets,
\bea
[x^n,p_m]_P&=&-\delta_m^n,
\nn\label{e181}\\
\{\theta_\ui^\alpha,\pi^\uj_\beta
\}_P&=&-\delta^\alpha_\beta\delta_\ui^\uj,\qquad
\{\bar\theta^{\ui\dot\alpha},\bar\pi_{\uj\dot\beta} \}_P
 = -\delta_{\dot\beta}^{\dot\alpha}\delta^\ui_\uj,
\nn\label{e182}\\
\{\theta_3^\alpha,\pi^3_\beta
\}_P&=&-\delta^\alpha_\beta,\qquad
\{\bar\theta^{3\dot\alpha},\bar\pi_{3\dot\beta} \}_P
 = -\delta_{\dot\beta}^{\dot\alpha},
\nn\label{e183}\\
{}[u^{+\ui},v^-_\uj]_P&=&-\delta^\ui_\uj,\qquad
[u^-_\ui,v^{+\uj}]_P=-\delta_\ui^\uj.
\label{e184}
\eea

The full list of constraints is given by
\bea
p^m p_m+R^{-2}D^{--}D^{++}-m^2&\approx&0,
\label{e185}\\
D^\ui_\alpha=-\pi^\ui_\alpha+ip_m(\sigma^m\bar\theta^\ui)_\alpha
 +z\theta^\ui_\alpha&\approx&0,
\label{e186}\\
\bar D_{\ui\dot\alpha}=\bar\pi_{\ui\dot\alpha}
-ip_m(\theta_\ui\sigma^m)_{\dot\alpha}-\bar z\bar\theta_{\ui\dot\alpha}
 &\approx&0,
\label{e187}\\
D^3_\alpha=-\pi^3_\alpha+ip_m(\sigma^m\bar\theta^3)_\alpha&\approx&0,
\label{e188}\\
\bar D_{3\dot\alpha}=\bar\pi_{3\dot\alpha}
-ip_m(\theta_i\sigma^m)_{\dot\alpha}&\approx&0,
\label{e189}\\
D^0-in&\approx&0,
\label{e190}\\
\chi_1=u^-_\ui u^{+\ui}-1&\approx&0,
\label{e191}\\
\chi_2=v^-_\ui u^{+\ui}+u^-_\ui v^{+\ui}&\approx&0.
\label{e192}
\eea
There is also one more extra harmonic constraint
\be
D^{++}\approx0,
\label{e193}
\ee
which is necessary to ``freeze'' the harmonic dynamics and to
keep only the physical degrees of freedom.

Constraints (\ref{e185},\ref{e190},\ref{e193}) are first-class and
should be imposed on the state upon quantization. Second-class
harmonic constraints (\ref{e191},\ref{e192}) are accounted by the
Dirac bracket,
\be
[f,g\}_D=[f,g \}_P +\frac12[f,\chi_2 ]_P[\chi_1,g
]_P -\frac12[f,\chi_1 ]_P[\chi_2,g ]_P.
\label{e196}
\ee
It is easy
to see that spinor constraints (\ref{e188},\ref{e189}) are
second-class, since we consider here the massive case. They can be
taken into account by using the Gupta--Bleuler method. In general,
if the mass of the superparticle is arbitrary and is not related
with the central charges $z$, $\bar z$, the spinor constraints
(\ref{e186},\ref{e187}) belong to the second class. The quantization
of such a particle model does not lead to physical supermultiplets.
Therefore, we consider  further
 only a special case, when the central charges are correlated
with the mass by BPS condition, \be z\bar z=m^2.
\label{e223}
\ee
In
this case, the superparticle model possesses the $\kappa$-symmetry
which is realized on the superspace coordinates as follows, \bea
\delta_\kappa\theta_{\ui\alpha}&=&
 -ip_m(\sigma^m\bar\kappa_\ui)_\alpha-\bar
 z\varepsilon_{\ui\uj}\kappa^\uj_\alpha,
\qquad
\delta_\kappa\bar\theta^\ui_{\dot\alpha}=
 ip_m(\kappa^\ui\sigma^m)_{\dot\alpha}+z\varepsilon^{\ui\uj}
  \bar\kappa_{\uj\dot\alpha},
\nn\label{e218}\\
\delta_\kappa x^m&=&i\delta_\kappa\theta_\ui\sigma^m\bar\theta^\ui
 -i\theta_\ui\sigma^m\delta_\kappa\bar\theta^\ui,
\qquad
\delta_\kappa e=-4(\bar\kappa_{\ui\dot\alpha}\dot{\bar\theta}{}^{\ui\dot\alpha}
 +\dot\theta_\ui^\alpha\kappa^\ui_\alpha),
\nn\label{e220}\\
\delta_\kappa\theta_{3\alpha}&=&0,\qquad
\delta_\kappa\bar\theta^3_{\dot\alpha}=0.
\label{e221}
\eea
The generators of $\kappa$-symmetry (\ref{e221}) given by
\be
\psi_{\ui\alpha}=ip_m\sigma^m_{\alpha\dot\alpha}\bar
D^{\dot\alpha}_\ui
 +\bar zD_{\ui\alpha}\approx0,\qquad
\bar\psi^\ui_{\dot\alpha}=-ip_m\sigma^m_{\alpha\dot\alpha}D^{\ui\alpha}
 -z\bar D^\ui_{\dot\alpha}\approx0,
\label{e224}
\ee
correspond to the additional first-class constraints.

\subsection{Gupta-Bleuler quantization}
Upon quantization, the canonical momenta are replaced
by the differential operators,
\bea
&&p_m\to i\frac\partial{\partial x^m},\quad
 \pi^\ui_\alpha\to-i\frac\partial{\partial\theta_\ui^\alpha},\quad
 \bar\pi_{\ui\dot\alpha}\to-i\frac\partial{\partial\bar\theta^{\ui\dot\alpha}},
 \nn\\&&
\pi^3_\alpha\to-i\frac\partial{\partial\theta_3^\alpha},\quad
 \bar\pi_{3\dot\alpha}\to-i\frac\partial{\partial\bar\theta^{3\dot\alpha}},
 \quad
 v^{+\ui}\to\frac\partial{\partial u^-_\ui},\quad
 v^-_\ui\to\frac\partial{\partial u^{+\ui}}.
\label{e199}
\eea
The spinor constraints (\ref{e186})--(\ref{e189}),
as well as the harmonic momenta (\ref{e177},\ref{e180}), turn into
the covariant spinor and harmonic derivatives, \bea
D^\pm_\alpha&=&u^\pm_\ui D^\ui_\alpha=
\pm\frac\partial{\partial\theta^{\mp\alpha}}
 +i(\sigma^m\bar\theta^\pm)_\alpha\partial_m-iz\theta^\pm_\alpha,
\label{a200}\\
\bar D^\pm_{\dot\alpha}&=&u^{\pm}_\ui\bar D^\ui_{\dot\alpha}=
\pm\frac\partial{\partial\bar\theta^{\mp\dot\alpha}}
 -i(\theta^\pm\sigma^m)_{\dot\alpha}\partial_m
 +i\bar z\bar\theta^\pm_{\dot\alpha},
\label{e202}\\
D^3_\alpha&=&\frac\partial{\partial\theta _3^\alpha}
 +i(\sigma^m\bar\theta^3)_\alpha\partial_m,
\qquad
\bar D_{3\dot\alpha}=-\frac\partial{\partial\bar\theta^{3\dot\alpha}}
 -i(\theta_3\sigma^m)_{\dot\alpha}\partial_m,
\label{e205}\\
D^{++}&=&u^+_\ui\frac\partial{\partial u^-_\ui},\qquad
D^{--}=u^-_\ui\frac\partial{\partial u^+_\ui},\qquad
D^0=u^{+\ui}\frac\partial{\partial u^{+\ui}}-u^{-\ui}\frac\partial{\partial
u^{-\ui}}
\label{e207}
\eea
with the following anticommutation relations
\bea
&&\{D^+_\alpha,D^-_\beta  \}=2iz\varepsilon_{\alpha\beta},\qquad
\{\bar D^+_{\dot\alpha},\bar D^-_{\dot\beta} \}=-2i\bar
z\varepsilon_{\dot\alpha\dot\beta},
\nn\label{e208}\\&&
\{D^+_\alpha,\bar D^-_{\dot\alpha}
\}=\{D^3_\alpha,\bar D_{3\dot\alpha} \}=-\{D^-_\alpha,\bar D^+_{\dot\alpha} \}=
-2i\sigma^m_{\alpha\dot\alpha}\partial_m,
\nn\label{e209}\\&&
\{D^+_\alpha, D^+_\beta \}=
\{D^-_\alpha, D^-_\beta \}=
\{\bar D^+_{\dot\alpha},\bar D^+_{\dot\beta}\}=
\{\bar D^-_{\dot\alpha},\bar D^-_{\dot\beta} \}=0.
\label{e210}
\eea

Supercharges (\ref{supercharges1}) in the harmonic superspace are
described now by the operators,
\bea
Q^\pm_\alpha&=&\mp\frac\partial{\partial\theta^{\mp\alpha}}
+i(\sigma^m\bar\theta^\pm)_\alpha\partial_m-iz\theta^\pm_\alpha,
\qquad \bar Q^\pm_{\dot\alpha}
 =\mp\frac\partial{\partial\bar\theta^{\mp\dot\alpha}}
 -i(\theta^\pm\sigma^m)_{\dot\alpha}\partial_m
  +i\bar z\bar\theta^\pm_{\dot\alpha},
\nn \label{Q2}\\
Q^3_\alpha&=&-\frac\partial{\partial\theta _3^\alpha}
 +i(\sigma^m\bar\theta^3)_\alpha\partial_m,
\qquad
\bar Q_{3\dot\alpha}=\frac\partial{\partial\bar\theta^{3\dot\alpha}}
 -i(\theta_3\sigma^m)_{\dot\alpha}\partial_m,
\label{Q5}
\eea
which form the $\cN=3$ superalgebra with a central
charge, \be \{
Q^+_\alpha,Q^-_\beta\}=-2iz\varepsilon_{\alpha\beta},\quad \{\bar
Q^+_{\dot\alpha},\bar Q^-_{\dot\beta} \}=2i\bar
z\varepsilon_{\dot\alpha\dot\beta},\quad \{Q^+_\alpha,\bar
Q^-_{\dot\alpha} \}=\{Q^3_\alpha,\bar Q_{3\dot\alpha}
\}=2i\sigma^m_{\alpha\dot\alpha}\partial_m. \label{N3-algebra1} \ee

Let us introduce the state $|\Phi\rangle=\Phi^{(n)}$, which is a
superfield on harmonic superspace with the equations of motion and
constraints originating from the superparticle constraints
(\ref{e185})--(\ref{e192}). The first-class constraint (\ref{e190})
leads to the following equation \be D^0\Phi^{(n)}=n\Phi^{(n)},
\label{e212} \ee which shows that this superfield has a definite
$U(1)$ charge. The other harmonic constraints
(\ref{e191},\ref{e192}) are accounted by the Dirac bracket
(\ref{e196}). The mass-shell constraint (\ref{e185}) is also
first-class, therefore we have \be
(\partial^m\partial_m-R^{-2}D^{--}D^{++}+m^2)\Phi^{(n)}=0.
\label{e213} \ee Furthermore, we require that the superfield
$\Phi^{(n)}$ obeys constraint (\ref{e193}), \be D^{++}\Phi^{(n)}=0,
\label{e214} \ee which removes all unphysical degrees of freedom.
Under this additional constraint (\ref{e214}) the mass-shell
condition (\ref{e213}) simplifies to \be
(\partial^m\partial_m+m^2)\Phi^{(n)}=0. \label{e215} \ee

Now we have to take into account the spinorial constraints
(\ref{e186})--(\ref{e189}) using Gupta-Bleuler method. The covariant
spinor derivatives should be divided into two subsets with commuting
constraints in each subset. Clearly, there are two ways of
separating these constraints into such subsets,
\bea
\{D^+_\alpha,\bar D^+_{\dot\alpha},\bar D_{3\dot\alpha} \}&\cup&
\{D^-_\alpha,\bar D^-_{\dot\alpha},D^3_\alpha \},
\label{e215.1}\\
\{D^+_\alpha,\bar D^+_{\dot\alpha},D^3_\alpha \}&\cup&
\{D^-_\alpha,\bar D^-_{\dot\alpha},\bar D_{3\dot\alpha} \}.
\label{e215.2}
\eea
Both these cases lead to equivalent results.
Therefore, we consider only (\ref{e215.1}) in detail. The
corresponding constraints \be D^+_\alpha\Phi^{(n)}=0,\quad \bar
D^+_{\dot\alpha}\Phi^{(n)}=0,\quad \bar D_{3\dot\alpha}\Phi^{(n)}=0,
\label{e216} \ee show that $\Phi^{(n)}$ is analytic with respect to
$\theta^+_\alpha$, $\bar\theta^+_{\dot\alpha}$ and is chiral in
$\theta_{3\alpha}$, $\bar\theta^3_{\dot\alpha}$ variables.

Finally, we have to take into account constraints (\ref{e224})
originating from the $\kappa$-sym\-met\-ry (\ref{e221}). Upon
quantization, the generators of $\kappa$-transformations
(\ref{e224}) turn into the differential operators \be
\psi^\pm_\alpha=-\sigma^m_{\alpha\dot\alpha}\partial_m\bar
D^{\pm\dot\alpha}
 +\bar z D^\pm_\alpha,\qquad
\bar\psi^\pm_{\dot\alpha}=\sigma^m_{\alpha\dot\alpha}\partial_m
D^{\pm\alpha}
 -z\bar D^\pm_{\dot\alpha}.
\label{e226} \ee
Note that owing to the analyticity (\ref{e216}),
constraints $\psi^+_\alpha\Phi^{(n)}=0$,
$\bar\psi^+_{\dot\alpha}\Phi^{(n)}=0$ are satisfied automatically,
while the ``$-$'' projections in (\ref{e226}) lead to the equations
\be (-\sigma^m_{\alpha\dot\alpha}\partial_m\bar D^{-\dot\alpha}
+\bar zD^-_\alpha)\Phi^{(n)}=0,\qquad
(\sigma^m_{\alpha\dot\alpha}\partial_m D^{-\alpha}-z\bar
D^-_{\dot\alpha}) \Phi^{(n)}=0. \label{e227} \ee

Let us introduce the operator
\be
Y^{--}=\frac i4(-z\bar D^-_{\dot\alpha}\bar D^{-\dot\alpha}
 +\bar z D^{-\alpha}D^-_\alpha),
\label{e228} \ee which commutes with covariant spinor derivatives as
\be \frac 1z[D^+_\alpha,Y^{--}]=\psi^-_\alpha,\qquad \frac1{\bar
z}[\bar D^+_{\dot\alpha},Y^{--}]= \bar\psi^-_{\dot\alpha}.
\label{e229} \ee As the superfield $\Phi^{(n)}$ is analytic, both
constraints (\ref{e227}) follow from
\be
(z\bar
D^-_{\dot\alpha}\bar D^{-\dot\alpha}
 -\bar zD^{-\alpha}D^-_\alpha)\Phi^{(n)}=0.
\label{e230} \ee
Despite this equation being stronger than the pair
(\ref{e227}), it should be imposed on the physical states as well,
since the first-class constraint (\ref{e228}) is a function of
spinorial constraints (\ref{e186},\ref{e187}) and forms the algebra
(\ref{e229}) with the generators of $\kappa$-symmetry. As a result,
all superparticle constraints are accounted by the corresponding
equations for the field $\Phi^{(n)}$.

\subsubsection{$\cN=3$ massive vector multiplet}
Let us consider a solution of constraints for the superfield
$\Phi^{(n)}$ on the physically interesting example of a superfield
$\Phi^{(1)}\equiv q^+$ with $U(1)$ charge $+1$. We will show that
such a superfield describes the $\cN=3$ massive vector multiplet.

To begin with, we list once again all the constraints for the $q^+$
superfield,
\bea
D^0 q^+&=&q^+,\label{e231.1}\\
D^+_\alpha q^+=\bar D^+_{\dot\alpha} q^+&=&0,\label{e231.2}\\
D^{++}q^+&=&0,\label{e231.3}\\
\bar D^3_{\dot\alpha}q^+&=&0,\label{e231.4}\\
{}[\bar z(D^-)^2 -z(\bar D^-)^2]q^+&=&0,\label{e231.5}\\
(\partial^m\partial_m+m^2)q^+&=&0.
\label{e231.6}
\eea
Equation
(\ref{e231.1}) is satisfied automatically, while the pair
(\ref{e231.5},\ref{e231.6}) follows from (\ref{e231.2},\ref{e231.3}).
To solve (\ref{e231.2}), we pass from the central coordinates to the
analytic ones,
\be
x^m_A=x^m-i(\theta^+\sigma^m\bar\theta^-+\theta^-\sigma^m\bar\theta^+),
\label{an-coord}
\ee
and transform Grassmann and harmonic
derivatives as well as the superfield $q^+$ such that $D^+_\alpha$,
$\bar D^+_{\dot\alpha}$ become short, \bea D^+_\alpha&\to&{\cal
D}^+_\alpha=e^\Omega D^+_\alpha e^{-\Omega}
=\frac\partial{\partial\theta^{-\alpha}},
\label{rot1}\\
\bar D^+_{\dot\alpha}&\to&\bar{\cal D}^+_{\dot\alpha}
 =e^\Omega \bar D^+_{\dot\alpha} e^{-\Omega}
 =\frac\partial{\partial\bar\theta^{-\dot\alpha}},
 \label{rot2}\\
D^-_\alpha&\to&{\cal D}^-_\alpha=e^\Omega D^-_\alpha e^{-\Omega}
=-\frac\partial{\partial \theta^{+\alpha}}
 +2i(\sigma^m\bar\theta^-)_\alpha\frac\partial{\partial x_A^m}-2iz\theta^-_\alpha,
\label{rot3}\\
\bar D^-_{\dot\alpha}&\to&\bar{\cal D}^-_{\dot\alpha}
=e^\Omega\bar D^-_{\dot\alpha} e^{-\Omega}
=-\frac\partial{\partial\bar\theta^{+\dot\alpha}}
-2i(\theta^-\sigma^m)_{\dot\alpha}\frac\partial{\partial x_A^m}
 +2i\bar z\bar\theta^-_{\dot\alpha},
\label{rot4}\\
D^{++}&\to&{\cal D}^{++}=e^\Omega D^{++} e^{-\Omega}
=D^{++}+iz(\theta^+)^2+i\bar z(\bar\theta^+)^2,
\label{rot5}\\
q^+&\to&{\bf q}^+=e^\Omega q^+, \eea where \be
\Omega=-iz\theta^+\theta^--i\bar z\bar\theta^+\bar\theta^-.
\label{omega} \ee In this representation, constraints (\ref{e231.2})
are solved automatically if ${\bf q}^+$ does not depend on
$\theta^-_\alpha$, $\bar\theta^-_{\dot\alpha}$, \be {\bf q}^+={\bf
q}^+(x^m_A,\theta^+_\alpha,\bar\theta^+_{\dot\alpha}
,\theta^3_\alpha,\bar\theta_{3\dot\alpha},u^\pm_\ui). \label{Qsup}
\ee Next, we expand (\ref{Qsup}) over $\theta^+_\alpha$,
$\bar\theta^+_{\dot\alpha}$ and solve the equation ${\cal
D}^{++}{\bf q}^+=0$ which follows from (\ref{e231.3}). As a result
we have \be {\bf q}^+=u^+_\ui F^\ui+\theta^{+\alpha}\Psi_\alpha
+\bar\theta^+_{\dot\alpha}\bar\Xi^{\dot\alpha} -iz(\theta^+)^2 F^\ui
u^-_\ui-i\bar z(\bar\theta^+)^2 F^\ui u^-_\ui
+2i\theta^+\sigma^m\bar\theta^+\partial_m F^\ui u^-_\ui,
\label{e232}
\ee
where all the components depend on
$(x^m_A,\theta_{3\alpha},\bar\theta^3_{\dot\alpha})$. Here $F^\ui$
is a doublet of complex scalars satisfying Klein-Gordon equation,
\be (\square+z\bar z)F^\ui=0, \label{e233} \ee and $\left(
\begin{smallmatrix} \Psi_\alpha \\
\bar\Xi^{\dot\alpha}\end{smallmatrix}
\right)$ is a massive Dirac spinor,
\be
i\sigma^{m\alpha\dot\alpha}\partial_m\Psi_\alpha-iz\bar\Xi^{\dot\alpha}=0,\qquad
i\sigma^m_{\alpha\dot\alpha}\partial_m\bar\Xi^{\dot\alpha}+i\bar
z\Psi_\alpha=0.
\label{e234}
\ee
In what follows, the mass is correlated with the central charges as
\be
m=iz=-i\bar z.
\label{BPS1}
\ee

Now we recall that ${\bf q}^+$ depends also on $\theta_{3\alpha}$,
$\bar \theta^3_{\dot\alpha}$ variables in a chiral way, \be \bar
D_{3\dot\alpha}{\bf q}^+=0.
\label{e235}
\ee
In the chiral
coordinates $y^m=x^m_A+i\theta_3\sigma^m\bar\theta^3$, the
derivative $\bar D_{3\dot\alpha}$ is short, $\bar
D_{3\dot\alpha}=-\frac\partial{\partial \bar\theta^{3\dot\alpha}}$,
and all components in (\ref{e232}) depend on $\theta_3^\alpha$, but
not on $\bar\theta^3_{\dot\alpha}$, \be
F^\ui=F^\ui(y^m,\theta_3^\alpha),\quad
\Psi_\alpha=\Psi_\alpha(y^m,\theta_3^\alpha),\quad
\bar\Xi^{\dot\alpha}=\bar\Xi^{\dot\alpha} (y^m,\theta_3^\alpha).
\label{e236} \ee Let us consider the decomposition of these
components in the series over $\theta_3^\alpha$, \bea
F^\ui(y,\theta_3)&=&f^\ui(y)+\theta_3^\alpha\sigma^\ui_\alpha(y)
+(\theta_3)^2 g^\ui(y),
\nn\label{e237}\\
\Psi^\alpha(y,\theta_3)&=&\psi^\alpha(y)+\theta_{3\beta}F^{(\alpha\beta)}(y)
+\theta_3^\alpha C+(\theta_3)^3 \lambda^\alpha(y),
\nn\label{e238}\\
\bar\Xi^{\dot\alpha}(y,\theta_3)&=&\bar \chi^{\dot\alpha}(y)
+\theta_{3\alpha}\bar A^{\alpha\dot\alpha}(y)
+(\theta_3)\bar\rho^{\dot\alpha}(y). \label{e239} \eea
Owing to
(\ref{e233}), the fields $f^\ui$, $g^\ui$, $\sigma^\ui_\alpha$ obey
the Klein-Gordon equation. Unfortunately, there is no Dirac equation
for the spinors $\sigma^\ui_\alpha$. However, the two Weyl spinors
$\sigma^\ui_\alpha$ with Klein-Gordon equation are equivalent to a
pair of Dirac spinors satisfying usual Dirac equation.\footnote{ Let
$\psi_\alpha$ be a function with the spinor index satisfying
Klein-Gordon equation, $\partial_m\partial^m
\psi_\alpha+m^2\psi_\alpha=0$. Factorizing the box operator we
rewrite this equation as $i\sigma^m_{\alpha\dot\alpha}\partial_m
\frac im\sigma^{n\beta\dot\alpha}\partial_n\psi_\beta
-m\psi_\alpha=0$. By denoting $\bar\chi^{\dot\alpha}=\frac im
\sigma^{m\alpha\dot\alpha}\partial_m\psi_\alpha$ the Klein-Gordon
equation can be rewritten as a pair
$i\sigma^m_{\alpha\dot\alpha}\partial_m\bar\chi^{\dot\alpha}-m\psi_\alpha=0$,
$i\sigma^m_{\alpha\dot\alpha}\psi^\alpha+m\bar\chi_{\dot\alpha}=0$
that is nothing but the Dirac equation for the spinor $\left(
\begin{smallmatrix}
\psi_\alpha\\\bar\chi^{\dot\alpha}
\end{smallmatrix} \right)$.}

Let us study the consequences of the Dirac equations (\ref{e234}).
They imply that
$\left(\begin{smallmatrix} \psi_\alpha \\ \bar\chi^{\dot\alpha}\end{smallmatrix}
\right)$ and $\left(\begin{smallmatrix} \lambda_\alpha \\ \bar\rho^{\dot\alpha}\end{smallmatrix}
\right)$ are usual Dirac spinors while the
components $\bar A_{\alpha\dot\alpha}$, $C$, $F_{\alpha\beta}$
obey
\bea
\sigma^m_{\alpha\dot\alpha}\partial_m C+\sigma^{m\beta}{}_{\dot\alpha}
 \partial_m F_{\alpha\beta}- z\bar A_{\alpha\dot\alpha}&=&0,
\label{e239.1}\\
\sigma^m{}_\alpha{}^{\dot\alpha}\partial_m\bar A_{\beta\dot\alpha}
+\bar z\varepsilon_{\alpha\beta} C+\bar z F_{\alpha\beta}&=&0.
\label{e239.2}
\eea
Equations (\ref{e239.1},\ref{e239.2}) have the following
solutions
\be
C=-\frac1{\bar z}\partial_m\bar A^m,\qquad
F_{\alpha\beta}=-\frac1{\bar z}\sigma^{mn}_{\alpha\beta}
 \bar F_{mn},
\label{e239.3}
\ee
\be (\square+z\bar z)\bar A_m=0,
\label{e239.4}
\ee
where $\bar F_{mn}=\partial_m\bar A_n-\partial_n\bar A_m$, $\bar
A_m=\frac12\sigma_m^{\alpha\dot\alpha}\bar A_{\alpha\dot\alpha}$ and
$\sigma^{mn}_{\alpha\beta}=-\frac14(\sigma^m_{\alpha\dot\alpha}\sigma^{n\dot\alpha}_\beta
-\sigma^n_{\alpha\dot\alpha}\sigma^{m\dot\alpha}_\beta)$. As a
result, vector $\bar A_n$ corresponds to the complex massive vector
field (with Proca equation) plus a complex scalar $C$ with
Klein-Gordon equation.

Summarizing these results we have the following field content in
${\bf q}^+$ subject to (\ref{e231.1})--(\ref{e231.6}):
\begin{itemize}
\item Complex massive vector $\bar A_m$ describes
6 bosonic degrees of freedom;
\item $f^\ui$, $g^\ui$, $C$ are complex scalars with 10 bosonic degrees of
freedom;
\item $\left(\begin{smallmatrix} \psi_\alpha \\ \bar\chi^{\dot\alpha}\end{smallmatrix}
\right)$, $\left(\begin{smallmatrix} \lambda_\alpha \\ \bar\rho^{\dot\alpha}\end{smallmatrix}
\right)$ are massive Dirac spinors with 8 fermionic degrees of
freedom;
\item The doublet of spinors $\sigma^\ui_\alpha$ does not satisfy
Dirac equation since it originates from the scalars $F^\ui$
obeying only the Klein-Gordon equation (\ref{e234}).
However, as noted above, $\sigma^\ui_\alpha$ correspond to two Dirac spinors
with 8 fermionic degrees of freedom.
\end{itemize}
This is nothing but the field content of $\cN=3$ massive vector
multiplet with BPS mass \cite{FS}. Note that it has double the
number of components in comparison with the massive (non-BPS)
$\cN=2$ vector multiplet.

\subsubsection{Superfield reduction of components in massive vector multiplet}
In the previous subsection, we have shown that the quantization of
the $\cN=3$ harmonic superparticle with central charges correlated
with the mass as in (\ref{BPS1}) leads to the $\cN=3$ massive vector
BPS multiplet. As this multiplet has double the  number of states in
comparison with the massive (non-BPS) $\cN=2$ vector multiplet, it
is natural to ask whether it is possible to impose additional
constraints on the $q^+$ superfield which reduce its component
content to the massive $\cN=2$ supergauge multiplet. As we will
show, this is possible if one relaxes the condition of the CPT
invariance of the multiplet.

To begin with, we introduce a conjugation ``$\smile$'' which acts as
a standard conjugation $\widetilde{\phantom{m}}$ (\ref{conj2}) in
harmonic superspace and changes the signs of the central charges
(and mass), $\breve z=-\bar z$, $\breve{\bar z}=-z$. For instance,
the superfield conjugated to (\ref{e232}) is \be \breve {\bf
q}^+=u^+_\ui \bar F^\ui-\theta^{+\alpha}\Xi_\alpha
+\bar\theta^+_{\dot\alpha}\bar\Psi^{\dot\alpha} -iz(\theta^+)^2\bar
F^\ui u^-_\ui-i\bar z(\bar\theta^+)^2\bar F^\ui u^-_\ui
+2i\theta^+\sigma^m\bar\theta^+\partial_m\bar F^\ui u^-_\ui.
\label{e240} \ee Note that ${\bf q}^+$ depends on
$\theta_{3\alpha}$, $\bar\theta^3_{\dot\alpha}$ in a chiral way
while $\breve {\bf q}^+$ is antichiral with respect to these
variables. It is natural to restrict the dependence on
$\theta_{3\alpha}$, $\bar\theta^3_{\dot\alpha}$ by imposing
equations which are similar to the ones in the massive Wess-Zumino
model, \be \frac14(D_3)^2{\bf q}^++iz\breve {\bf q}^+=0,\qquad
\frac14(\bar D^3)^2\breve {\bf q}^+-i\bar z {\bf q}^+=0.
\label{e241}
\ee
Equations (\ref{e241}) are conjugate to each other
with respect to $\smile$ conjugation and have the following
consequences: \bea \frac14(D^3)^2 F^\ui+iz\bar F^\ui=0,&\quad&
 \frac14(\bar D_3)^2\bar F^\ui-i\bar z F^\ui=0,
\label{e242}\\
\frac14(D^3)^2\Psi_\alpha-iz\Xi_\alpha=0,&&
 \frac14(\bar D_3)^2\Xi_\alpha+i\bar z \Psi_\alpha=0,
\label{e243}\\
\frac14(D^3)^2\bar \Xi^{\dot\alpha}+iz \bar \Psi^{\dot\alpha}=0,&&
\frac14(\bar D_3)^2\bar \Psi^{\dot\alpha}-i\bar z\bar
\Xi^{\dot\alpha}=0.
\label{e244}
\eea
Clearly, (\ref{e242}) are
nothing but the usual Wess-Zumino equations for the $\cN=1$
superfields $F^\ui$, $\bar F^\ui$. Therefore for the components of
these superfields we have \bea &&(\square+z\bar z)f^\ui=0,\qquad
 (\square+z\bar z)\bar f^\ui=0,
\label{e245}\\
&&i\sigma^m_{\alpha\dot\alpha}\partial_m\bar\sigma^{\ui\dot\alpha}
 +i\bar z\sigma^\ui_\alpha=0,\qquad
i\sigma^m_{\alpha\dot\alpha}\partial_m\sigma^{\ui\alpha}
 +iz\bar \sigma^\ui_{\dot\alpha}=0,
\label{e246}\\
&&g^\ui=iz\bar f^\ui,\qquad
 \bar g^\ui=-i\bar z f^\ui.
 \label{e246.1}
\eea
One can easily construct a Dirac spinor
$\left(\begin{smallmatrix}
\sigma^2_\alpha \\\bar\sigma^{2\dot\alpha}
\end{smallmatrix}\right)\equiv\left(\begin{smallmatrix}
\rho_\alpha \\\bar \mu^{\dot\alpha}
\end{smallmatrix}\right)$
from the spinors $\sigma^\ui_\alpha$ satisfying (\ref{e246}).
Note that the conjugated spinor
$\left(\begin{smallmatrix}
\mu_\alpha \\\bar \rho^{\dot\alpha}
\end{smallmatrix}\right)=\left(\begin{smallmatrix}
\sigma^1_\alpha \\-\bar\sigma^{1\dot\alpha}
\end{smallmatrix}\right)$ satisfies the Dirac equation with opposite
sign of the mass.

Let us consider the pair of equations (\ref{e243}). They lead to the
Klein-Gordon equations for spinors $\chi_\alpha$, $\psi_\alpha$,
while $\lambda_\alpha$, $\rho_\alpha$ are expressed from them, \be
\lambda_\alpha=-iz\chi_\alpha,\qquad \rho_\alpha=iz\psi_\alpha.
\label{e246.2} \ee The other components obey the following equations
\bea \sigma^m{}_\beta{}^{\dot\alpha}\partial_m A_{\alpha\dot\alpha}
+\bar z\varepsilon_{\alpha\beta} C+\bar zF_{\alpha\beta}&=&0,
\label{e247}\\
\sigma^{m\beta}{}_{\dot\beta}\partial_m F_{\alpha\beta}
+\sigma^m_{\alpha\dot\beta}\partial_m C+z A_{\alpha\dot\beta}&=&0,
\label{e248} \eea which are solved by \be C=\frac1{\bar z}\partial_m
A^m,\qquad F_{\alpha\beta}=-\frac1{\bar z}\sigma^{mn}_{\alpha\beta}
F_{mn}, \label{e249} \ee \be (\square+z\bar z)A_m=0.
\label{e250}
\ee
Considering (\ref{e249}) together with (\ref{e239.3}) we
conclude that field strength  $F_{mn}=\partial_m A_n-\partial_n A_m$
is real, $F_{mn}=\bar F_{mn}$, while $\partial_m A^m$ is imaginary
and corresponds to a real scalar $B=i\partial_m A^m$. As a result,
the complex vector $A_n$ splits into a real vector obeying Proca
equations, and a real scalar with Klein-Gordon equation. The
resulting multiplet exactly corresponds to a massive $\cN=2$ vector
multiplet:
\begin{itemize}
\item Two complex scalars $f^i$ and the real one $B$ give 5 real bosonic
degrees of freedom;
\item Dirac spinors $\left(\begin{smallmatrix} \psi_\alpha \\ \bar\chi^{\dot\alpha}\end{smallmatrix}
\right)$, $\left(\begin{smallmatrix}
\rho_\alpha \\\bar \mu^{\dot\alpha}
\end{smallmatrix}\right)$ describe 8 fermionic degrees of freedom;
\item Real massive vector field $A_m$ has 3 bosonic components on-shell.
\end{itemize}

To show that equations (\ref{e241}) preserve the $\cN=3$
supersymmetry with a central charge, we note that the supercharges
(\ref{Q5}) are conjugate to each other with respect to $\smile$
conjugation rather than to $\widetilde{\phantom{a}}$, \be
\stackrel{\smile}{Q^+_\alpha}=-\bar Q^+_{\dot\alpha},\quad
\stackrel{\smile}{Q^-_\alpha}=-\bar Q^-_{\dot\alpha},\quad
\stackrel{\smile}{\bar Q^+_{\dot\alpha}}=Q^+_\alpha,\quad
\stackrel{\smile}{\bar Q^-_{\dot\alpha}}=Q^-_\alpha. \label{Qconj}
\ee Hence, the supersymmetry variation is real under such a
conjugation, \be \delta_{\epsilon}=-\epsilon^{+\alpha}Q^-_\alpha
+\bar\epsilon^+_{\dot\alpha}\bar Q^{-\dot\alpha}
+\epsilon^{-\alpha}Q^+_\alpha -\bar\epsilon^-_{\dot\alpha}\bar
Q^{+\dot\alpha} +\epsilon_3^\alpha Q^3_\alpha
+\bar\epsilon^3_{\dot\alpha} \bar Q_3^{\dot\alpha}
=\,\stackrel{\smile}{\delta_\epsilon}. \label{susy-real} \ee
Therefore, both superfields ${\bf q}^+$ and $\breve {\bf q}^+$
transform in the same way under supersymmetry, and the conjugation
$\smile$ in equations (\ref{e241}) does not break the $\cN=3$
supersymmetry. However, the resulting multiplet is not CPT
selfconjugated and the CPT symmetry is lost. In terms of superfields
it is obvious since the conjugation $\smile$ involves the change of
the sign of the mass and the central charge. In components it leads
to the
fact that the spinors $\left(\begin{smallmatrix} \psi_\alpha \\
\bar\chi^{\dot\alpha}\end{smallmatrix} \right)$,
$\left(\begin{smallmatrix} \rho_\alpha \\\bar \mu^{\dot\alpha}
\end{smallmatrix}\right)$ are standard Dirac ones while their conjugates
$\left(\begin{smallmatrix} \chi_\alpha \\ \bar\psi^{\dot\alpha}\end{smallmatrix}
\right)$, $\left(\begin{smallmatrix}
\mu_\alpha \\\bar \rho^{\dot\alpha}
\end{smallmatrix}\right)$ satisfy the Dirac equation with opposite sign of
the mass.

As a result, equations (\ref{e241}) reduce the number of components
in the $\cN=3$ vector multiplet by half resulting in the $\cN=2$
massive (non-BPS) vector multiplet. In other words, the $\cN=2$
massive vector multiplet is equivalent to a half of the $\cN=3$
vector multiplet which respects the $\cN=3$ supersymmetry with
central charge, but is not CPT selfconjugated. Although this fact is
well known \cite{FS}, we establish this correspondence by superfield
considerations. In particular, this multiplet is realized as a
single constrained $\cN=3$ superfield. It would also be very
tempting to find a supersymmetric action for superfields ${\bf
q}^+$, $\breve {\bf q}^+$ reproducing the corresponding equations of
motion.

\section{Conclusion}

Let us summarize the results obtained in the quantization
of $\cN=3$ superparticle.
\begin{enumerate}
\item The models of the $\cN=3$ superparticles both with and without
central charge term, are considered in the harmonic superspace.
Since the $\cN=3$ superalgebra with central charge possesses
$SU(2)\times U(1)$ R-symmetry rather than $U(3)$, the description of
the $\cN=3$ superparticle with central charge is achieved in the
$\cN=2$ harmonic superspace (with $SU(2)$ harmonic variables)
extended by a pair of extra Grassmann variables:
$\{x^m,\theta^\pm_\alpha,\bar\theta^\pm_{\dot\alpha},
\theta_{3\alpha},\bar\theta^3_{\dot\alpha},u^\pm_\ui \}$.
\item By quantizing the $\cN=3$ superparticle without central
charge, we obtain $\cN=3$ superfield realizations of the $\cN=3$
supergauge multiplet and the gravitino multiplet (with highest
helicity 3/2). The latter is described by a chiral $\cN=3$
superfield satisfying linearity constraints, while the former is
given by the superfield strengths, which are short superfields in
the $\cN=3$ harmonic superspace subject to Grassmann and harmonic
shortness conditions. These superfields were originally introduced
in \cite{Sorokin1} and studied in \cite{Ferrara}.
\item The $\cN=3$ superparticle with central charge is quantized
similarly to the $\cN=2$ superparticle in harmonic superspace
\cite{Sorokin1,Sorokin2}. The resulting massive $\cN=3$ supergauge
multiplet is given by 5 complex scalars, 4 Dirac spinors and 1
complex vector on-shell. It is embedded into a superfield $q^+$,
which is analytic in $\theta^\pm_\alpha,\bar\theta^\pm_{\dot\alpha}
$ variables and chiral with respect to
$\theta_{3\alpha},\bar\theta^3_{\dot\alpha}$. The equation of motion
for this superfield is similar to the one for $q$-hypermultiplet,
$D^{++}q^+$=0.
\item We notice that the number of states of the massive $\cN=3$
supergauge multiplet is doubled compared to the massive (non-BPS)
$\cN=2$ vector multiplet with 5 real scalars 4 spinors and 1 real
vector. The doubling of states in the representations of the $\cN=3$
superalgebra with central charge is required for CPT invariance
\cite{FS}. However, if we abandon the CPT invariance, the numbers of
states in these two multiplets coincide. We have shown how this can
be achieved at a superfield level: by imposing the extra superfield
constraints on the $\cN=3$ superfield $q^+$ we reduce the number of
states by one half, arriving at the massive $\cN=2$ supergauge
multiplet realized as a constrained $\cN=3$ superfield. Of course,
these constraints break the CPT invariance manifestly. In
components, the loss of CPT invariance means that the Dirac spinor
$\left(\begin{smallmatrix} \lambda_\alpha\\\bar\mu^{\dot\alpha}
\end{smallmatrix} \right)$
and its conjugate $\left(\begin{smallmatrix}
\mu_\alpha\\\bar\lambda^{\dot\alpha} \end{smallmatrix}
\right)$ satisfy Dirac equations with different signs of the mass.
\end{enumerate}

The results of the quantization of the massless $\cN=3$
superparticle without a central charge term are rather expected: the
superfield realizations of the $\cN=3$ supergauge and gravitino
multiplets are achieved. However, the quantization of the $\cN=3$
superparticle with a central charge leads to a superfield
description for the massive $\cN=3$ supergauge multiplet which was
previously unknown. It would be interesting to develop the classical
and quantum field theory of this multiplet in a superfield
realization. If a Lagrangian superfield formulation of this model is
achievable, the classical and quantum properties would be as
interesting as for the massive $\cN=2$ vector superfield
\cite{G,K,Pletnev}.

We have shown the relations between the $\cN=3$ and the $\cN=2$
massive vector multiplets to be even deeper. Indeed, the $\cN=2$
massive vector multiplet is described by an $\cN=3$ superfield under
specific superfield constraints which manifestly break CPT symmetry
of $\cN=3$ superalgebra with central charge. In other words, the
$\cN=2$ massive vector multiplet can be viewed as a half of the
$\cN=3$ massive vector multiplet, which does not have its CPT
conjugate. This means that the free $\cN=2$ massive vector multiplet
possesses $\cN=3$ supersymmetry with a central charge, if we neglect
CPT invariance. Finally, the $\cN=3$ superfield under the additional
constraints can be considered as an alternative formulation for the
massive $\cN=2$ vector multiplet. It would be interesting to study
whether it is possible to include the non-Abelian selfinteraction of
this multiplet in terms of $\cN=3$ superfields, and to build some
action directly in the $\cN=3$ superspace.

Another obvious continuation of this research would be the study of
the $\cN=4$ superparticle in a similar way. The case of the massive
superparticle with a central charge would be the most tempting and
should lead to a massive $\cN=4$ vector multiplet realized as an
$\cN=4$ superfield. This model would be of high interest both at the
classical and at the quantum level.

\vspace{5mm} {\bf Acknowledgements.} The authors are grateful to
O. Lechtenfeld, A.V. Galajinsky, N.G. Pletnev, D. Sorokin, B.M.
Zupnik for stimulating discussions. We would like to thank also J.
Br\"odel for careful reading and improving the text. I.B.S. is
very grateful to D. Sorokin for intensive consultations on the
aspects of superparticles quantization and for critical reading of
the draft and to INFN, Sezione di Padova {\rm\&} Dipartimento di
Fisica ``Galileo Galilei", Universit\`{a} degli Studi di Padova
for kind hospitality and support during the period when the
essential part of this work was done. The present work is
supported particularly by INTAS grant, project No. 05-1000008-7928,
by RFBR grants, projects No. 06-02-16346 and No. 08-02-90490, by DFG
grant, project No 436 RUS/113/669/0-3 and grant for LRSS, project
No. 2553.2008.2. I.B.S. acknowledges the support from INTAS grant,
project No. 06-1000016-6108.

\end{document}